\documentclass[aps,pra,prabib,amsmath,amssymb,amsfonts,
epsfig,smallcaptions,onecolumn,showpacs,superscriptaddress]{revtex4}
\usepackage{psfig}
\usepackage{graphicx}

\newcommand{\eqname}[1]{\label{eq:#1}}
\newcommand{\bgar}{\begin{eqnarray}}
\newcommand{\enar}[1]{\label{eq:#1}\end{eqnarray}}

\newcommand{\ket}[1]{ | #1 \rangle }

\newcommand{\braket}[2]{ \langle #1 | #2 \rangle }
\newcommand{\braopket}[3]{ \left\langle #1 \right| #2 \left| #3 \right\rangle }
\newcommand{\ketbra}[2]{ | #1 \left\rangle \right\langle #2 |}

\newcommand{\eq}[1]{(\ref{eq:#1})}
\newcommand{\al}[1]{^{(#1)}}
\newcommand{\Psihd}{\hat\Psi^\dagger}

\newcommand{\Psih}{\hat\Psi}

\newcommand{\ahd}{\hat a^\dagger}
\newcommand{\ah}{\hat a}

\newcommand{\chd}{\hat c^\dagger}
\newcommand{\ch}{\hat c}
\newcommand{\updown}{{\uparrow \downarrow}}
\newcommand{\upa}{\uparrow}
\newcommand{\doa}{\downarrow}

\begin{document}

\title{Coherence and correlation properties of a one-dimensional
  attractive Fermi gas}

\affiliation{Laboratoire Kastler Brossel, \'Ecole Normale
Sup\'erieure, 24 rue Lhomond, 75231 Paris Cedex 05, France}
\affiliation{CRS BEC-INFM and Dipartimento di Fisica, Universit\`a di
  Trento, I-38050 Povo, Italy}

\author{Iacopo Carusotto}
\affiliation{Laboratoire Kastler Brossel, \'Ecole Normale
Sup\'erieure, 24 rue Lhomond, 75231 Paris Cedex 05, France}
\affiliation{CRS BEC-INFM and Dipartimento di Fisica, Universit\`a di
  Trento, I-38050 Povo, Italy}

\author{Yvan Castin}
\email{Yvan.Castin@lkb.ens.fr}
\affiliation{Laboratoire Kastler Brossel, \'Ecole Normale
Sup\'erieure, 24 rue Lhomond, 75231 Paris Cedex 05, France}

\begin{abstract}
A recently developed Quantum Monte Carlo algorithm based on
the stochastic evolution of Hartree-Fock states has been applied to compute
the static 
correlation functions of a one-dimensional model of attractively interacting
two component fermions. The numerical results have been extensively compared to existing
approximate approaches.
The crossover to a condensate of pairs can be identified as the first-order
pair coherence extending throughout the whole size of the system. The
possibility of revealing the onset of the transition with other observables
such as the density-density correlations or the second-order momentum space
correlations is discussed. 
\end{abstract}


\pacs{05.30.Fk, 02.70.Ss  }

\date{\today}

\maketitle

\section{Introduction}


The recent developments in the cooling and trapping techniques of neutral
atoms have opened the way to the realization of fermionic atomic samples at
temperatures well below the degeneracy temperature~\cite{FermiDegen}.
This suggests that atomic gases are ideal candidates for the study
of the physics of degenerate many-fermion systems.
With respect to solid state ones, atomic systems offer in
fact a better isolation from external disturbances such as material defects, a
better knowledge of the microscopic details of the systems, as well as a wider
range of tunability of the parameters, in particular the interparticle
interactions.
By tuning the external magnetic field around a Feshbach
resonance, the atom-atom scattering length $a$ can be varied
from $k_F a=-\infty$ to $+\infty$ ($k_F$ being the Fermi momentum)
opening the way towards a 
comprehensive study of the pairing transition both in the 
regime $a>0$ in which a Bose-condensate (BEC) of tightly-bound
molecules is present, and in the regime $a<0$ (BCS) in
which a condensate of Cooper pairs is formed. 
Diatomic molecules have been created and observed by several
experimental groups~\cite{Molecules}. Bose-Einstein condensation of 
tightly bound diatomic molecules has been recently
reported~\cite{MolecBEC}. The crossover region between BEC and BCS is
currently under experimental investigation~\cite{CrossoverExp} and
first evidences of pairing in the crossover region have been reported
in~\cite{AtomicBCS}.

From the theoretical point of view, a large effort is currently made to
establish the main features of the pairing for high values of the
scattering length $k_F |a|\gg 1$, regime in which the atomic gas shows
strong correlations~\cite{NSR,Randeria,CrossoverTheory}. 
In particular, the dependance of the transition temperature on the interaction
strength in this crossover region is still an open problem.

The present paper reports a numerical study of the condensation of
pairs in a
regime of relatively strong interactions, so to characterize the consequences
of the transition on the different observables of the system and
identify specific features which may represent unambiguous signatures of the
onset of condensation of pairs.

The calculations have been performed by applying the quantum Monte Carlo (QMC)
method developed in~\cite{Chomaz} to a one-dimensional lattice model
of fermions with attractive on-site interactions. A short description
of the model under examination is given in sec.~\ref{sec:model}, while
the numerical algorithm used for the calculations is presented in
sec.~\ref{sec:QMC}.  
Numerical results are presented in sec.~\ref{sec:results} and then extensively
compared to the predictions of a perturbative expansion in the
interaction coupling constant (sec.~\ref{sec:Pert}), and of existing
approximate approaches (sec.\ref{sec:Exist}), such as  the BCS
theory~\cite{LandauCM,deGennes}, two versions of the random phase
approximation (RPA)~\cite{FetterWalecka,Mahan} as well  
as the Nozi\`eres Schmitt-Rink theory~\cite{NSR}.

Several among the most relevant correlation functions of the Fermi gas have
been  considered here, in particular the opposite-spin density-density correlation 
function $\langle \rho_\downarrow(x)\,\rho_\uparrow(0)\rangle$, the first-order
pair coherence function $\langle \Psihd_\doa(x)\,\Psihd_\upa(x)\,
\Psih_\upa(0)\,\Psih_\doa(0) \rangle$ and the second-order momentum space correlation
function $\big\langle {\hat n}_{k\upa}\,{\hat n}_{-k\doa}\big\rangle$. 
The density-density correlation
function has been already the object of several papers studying the
experimental signatures of the BCS transition in atomic Fermi systems,
e.g.~\cite{g2ud_obs}, while the first-order
pair coherence function is the counterpart, in a
non-symmetry-breaking approach,  of the order parameter of the phase
transition in a Landau-Ginzburg theory~\cite{LandauCM}.

\section{The physical system}
\label{sec:model}

A one-dimensional low energy two-component Fermi gas can be modeled by
the Hamiltonian:
\begin{equation}
{\mathcal H}=\sum_{k,\sigma} \frac{\hbar^2 k^2}{2m} \ahd_{k\sigma}
\ah_{k\sigma} + 
g_0\sum_x dx\,\Psihd_\uparrow(x)\Psihd_{\downarrow}(x)
\Psih_{\downarrow}(x)\Psih_\uparrow(x).
\eqname{Hamilt}
\end{equation}
The spatial coordinate $x$ runs on a discrete lattice of ${\mathcal
 N}$ points with periodic boundary conditions; $L$ is the total length
 of the quantization box and $dx=L/{\mathcal N}$ is the length
of the unit cell of the lattice. The spin index runs over the two
 $\sigma=\uparrow,\downarrow$ spin  states. 
The system is taken as spatially homogeneous, $m$ is the atomic mass,
 and interactions are modeled by a two-body discrete delta potential
 with a coupling constant $g_0$. 
The field operators $\Psih_\sigma(x)$ satisfy the usual fermionic
anticommutation relations
 $\{\Psih_\sigma(x),\Psihd_{\sigma'}(x')\}=\delta_{\sigma,\sigma'}
\,\delta_{x,x'}/dx$  and can be expanded on plane waves according to
 $\Psih_\sigma(x)=\sum_{k}\ah_{k\sigma}e^{ikx}/\sqrt{L}$ with
 $k$ restricted to the first Brillouin zone of the reciprocal
 lattice. In order for the discrete model to correctly reproduce the
 underlying continuous field theory, the grid spacing $dx$ must be smaller
 than all the relevant length scales of the system, e.g. the thermal
 wavelength and the mean interparticle spacing.
In the present one-dimensional case, the
 relation between the coupling  constant on the  lattice and the
 physical 1D coupling constant $g_{1D}$ is:
\begin{equation}
g_0=g_{1D}\, \Big(1+\frac{m\,g_{1D}\,dx}{\pi^2\, \hbar^2}\Big)^{-1},
\end{equation}
which, in the limit $dx\ll \pi^2\, \hbar^2/m\,g_{1D}$ reduces to the expected one
$g_0=g_{1D}$~\cite{MoraThese,YvanHouches}.
This condition is satisfied in the Monte Carlo simulations presented
 in this paper.
We also note that two particles interacting in free space with a attractive delta potential
in 1D have a bound state of energy $-m g_{1D}^2/4\hbar^2$. In the numerical examples of this
paper, the Fermi energy is much larger than this binding energy so that we are not investigating
the condensation of preformed pairs but rather a BCS regime.

\section{The Quantum Monte Carlo scheme}
\label{sec:QMC}

We assume the gas to be at thermal equilibrium at a temperature $T$
in the canonical ensemble, so that the unnormalized density operator
$\rho_{\rm eq}(\beta)=e^{-\beta {\mathcal H}}$ with $\beta=1/k_B T$.
From textbook statistical physics, we know that such a density
operator can be obtained by means of an imaginary-time evolution:
\begin{equation}
\frac{d\rho_{\rm eq}(\tau)}{d\tau}=-\frac{1}{2}[{\mathcal H}\rho_{\rm
    eq}(\tau)+\rho_{\rm eq}(\tau){\mathcal H}]
\eqname{ImagTev}
\end{equation}
during a ``time'' interval $\tau=0\rightarrow \beta$ starting from the
initial state corresponding to the infinite temperature case
where $\rho_{\rm eq}(\tau=0)={\mathbf 1}_N$, ${\mathbf 1}$ being the
identity matrix in the $N$-body Hilbert space.

As it has been recently shown in~\cite{Chomaz}, the exact solution of the
imaginary-time evolution \eq{ImagTev} can be written as a statistical
average of Hartree-Fock dyadics of the form:
\begin{equation}
\sigma=\ketbra{\phi\al{1}_1\ldots\phi\al{1}_N}
{\phi\al{2}_1\ldots\phi\al{2}_N}.
\eqname{HFAnsatz}
\end{equation}
For $\alpha=1,2$, $\phi\al{\alpha}_j$ ($j=1\ldots N$) are
Hartree-Fock orbitals for the $N$ fermions, in the sense that:
\begin{equation}
\ket{\phi\al{\alpha}_1\ldots\phi\al{\alpha}_N}=
\ahd_{\phi_1^{(\alpha)}}\ldots\ahd_{\phi_N^{(\alpha)}}\,\ket{0},
\eqname{HFAnsatz2}
\end{equation}
the creation operator corresponding to the wavefunction
$\phi(x,\sigma)$ being defined as:
\begin{equation}
\ahd_{\phi}=\sum_{x,\sigma}\,dx\,\phi(x,\sigma)\,\Psihd_\sigma(x).
\eqname{HFAnsatz3}
\end{equation}
For the model Hamiltonian \eq{Hamilt}, the imaginary-time evolution of
each of the orbitals $\phi\al{\alpha}_j$ can be reformulated in terms
of Ito stochastic differential equations of the form:
\begin{multline}
\eqname{Ito_phi}
d\phi\al{\alpha}_i(x,\sigma)=
-\frac{d\tau}{2}\left\{
\frac{P^2}{2m}\phi_i\al{\alpha}(x,\sigma)+ \right.\\
+g_0
\sum_j\frac{1}{\|\phi\al{\alpha}_j\|^2}\,\Big[
|\phi_j\al{\alpha}(x,-\sigma)|^2
\phi_i\al{\alpha}(x,\sigma)-\phi^{(\alpha)*}_j(x,-\sigma)\,
\phi\al{\alpha}_j(x,\sigma)\,\phi\al{\alpha}_i(x,-\sigma)\Big]+ \\
-\frac{g_0}{2}\sum_j\sum_{x',\sigma'}dx'\,
\frac{1}{\|\phi_j\al{\alpha}\|^2\,\|\phi_i\al{\alpha}\|^2}\,
\Big[
\phi_i^{(\alpha)*}(x',\sigma')\,\phi_j^{(\alpha)*}(x',-\sigma')\,
\phi_j\al{\alpha}(x',-\sigma')\,\phi_i\al{\alpha}(x',\sigma')+ \\
\left.
 -\phi_i^{(\alpha)*}(x',\sigma')\,\phi_j^{(\alpha)*}(x',-\sigma')\,
\phi_j\al{\alpha}(x',\sigma')\,\phi_i\al{\alpha}(x',-\sigma')
\Big]\,\phi_i\al{\alpha}(x,\sigma)\right\}+dB_i\al{\alpha}(x,\sigma),
\end{multline}
where $P$ represents the momentum operator on the grid and the norm
$\|\phi\|$ is defined as $\|\phi\|^2=\sum_{x\sigma}
dx\,|\phi(x,\sigma)|^2$. 
The deterministic part is simply the mean-field Hartree-Fock
equation in imaginary time, 
while the correlation functions of the zero-mean noise
$dB_i\al{\alpha}$ are given by:
\begin{equation}
\overline{dB\al{\alpha}_i(x,\sigma)\,dB\al{\alpha'}_j(x',\sigma')}=
-\frac{g_0}{2\,dx}\,d\tau\,{\mathcal Q}\al{\alpha}_{\perp\,(x,\sigma)}
\,{\mathcal Q}\al{\alpha'}_{\perp\,(x',\sigma')}\,
\left[\phi\al{\alpha}_i(x,\sigma)\,\phi\al{\alpha'}_j(x',\sigma')
\,\delta_{\alpha,\alpha'}\,\delta_{\sigma,-\sigma'}
\,\delta_{x,x'}\right]. 
\eqname{noise}
\end{equation}
The projector ${\mathcal Q}\al{\alpha}_{\perp\,(x,\sigma)}$ projects
orthogonally to the subspace spanned by the wavefunctions
$\phi\al{\alpha}_j(x,\sigma)$. 
A possible noise with the required
correlation function \eq{noise} is: 
\begin{equation}
\left(
\begin{array}{c}
dB_i\al{\alpha}(x,\uparrow) \\
dB_i\al{\alpha}(x,\downarrow)
\end{array}
\right)
=
\sqrt{-\frac{g_0}{2\,dx}d\tau}
{\mathcal Q}\al{\alpha}_{\perp}
\left(
\begin{array}{cc}
\xi^{(\alpha)}(x) & 0 \\
0 & \xi^{(\alpha)*}(x) \\ 
\end{array}
\right)
\left(
\begin{array}{c}
\phi_i\al{\alpha}(x,\uparrow) \\
\phi_i\al{\alpha}(x,\downarrow)
\end{array}
\right)
,
\eqname{noise2}
\end{equation}
with $\xi^{(\alpha)}(x)$ independent zero-mean Gaussian noises with
$\overline{\xi^{(\alpha)}(x)\,\xi^{(\alpha')}(x')}=0$, 
$\overline{\xi^{(\alpha)*}(x)\,\xi^{(\alpha')}(x')}=\delta_{x,x'}
\delta_{\alpha,\alpha'}$.
It can be proven \cite{Chomaz} that this set of stochastic differential equations
reproduces, in the average over the noise, the exact evolution of the Hartree-Fock
dyadic $\sigma$ during $d\tau$:
\begin{equation}
\overline{d\sigma}=
-\frac{d\tau}{2}\big[{\mathcal H}\sigma+\sigma{\mathcal H}\big].
\end{equation}
The initial state ${\mathbf 1}_N$ can be written as a functional
integral over all possible sets of orthonormal wavefunctions
$\{\phi_j^{(0)}(x,\sigma)\}$ ($j=1\ldots N$):
\begin{equation}
{\mathbf 1}_N=\int_1\,{\mathcal D}\phi\al{0}_1\ldots{\mathcal D}\phi\al{0}_N\,
\ketbra{\phi\al{0}_1\ldots\phi\al{0}_N}{\phi\al{0}_1\ldots\phi\al{0}_N}.
\eqname{identity}
\end{equation}
This writing of the identity operator can be used as a starting point for an
exact simulation of the fermionic many-body problem.  
To this purpose, we have to numerically solve the stochastic
differential equations \eq{Ito_phi} for imaginary times going from
$\tau=0$ to $\tau=\beta$. This is done by splitting the
imaginary-time interval into a large enough number ${\mathcal M}$ of
time steps; $\xi^{(\alpha)}_j(x)$ is the noise terms at the time-step $j$
($j=1\ldots {\mathcal M}$) on the site $x$.
The expectation values of any
observable at temperature $T$ is then obtained as an average over all
the possible values of the initial wavefunctions $\phi\al{0}_i$ and
the elementary noise terms $\xi^{(\alpha)}_j(x)$.

For example, the partition function $\textrm{Tr}[\rho]$ is obtained as:
\begin{equation}
\textrm{Tr}[\rho]=\overline{ \braket{\phi\al{2}_1\ldots\phi\al{2}_N}
{\phi\al{1}_1\ldots\phi\al{1}_N} }
\end{equation}
or, equivalently, as the determinant $\textrm{Det}[M]$ of the matrix
$M$ whose entries are $M_{ij}=\braket{\phi\al{2}_j}{\phi\al{1}_i}$.
In the following, we shall be mainly interested in the one- and two-body
correlation functions of the gas.
By making use of the Jacobi theorem~\cite{Jacobi}, these can
be usefully rewritten in the 
following compact forms:
\begin{equation} 
\Big\langle\Psihd_\sigma(x)\Psih_{\sigma'}(x')\Big\rangle=
\textrm{Det}[M]\,\sum_{ij}\,\big(M^{-1}\big)_{ij}\,\phi^{(2)*}_i(x,\sigma)\,
\phi^{(1)}_j(x',\sigma')
\end{equation}
and
\begin{multline}
\Big\langle\Psihd_\sigma(x)\,\Psihd_{\sigma'}(x')
\Psih_{\sigma''}(x'')\Psih_{\sigma'''}(x''')\Big\rangle=\\
=\textrm{Det}[M]\,\sum_{ijkl}
\,\textrm{Det}\left[
\begin{array}{cc}
\big(M^{-1}\big)_{il} & \big(M^{-1}\big)_{ik} \\
\big(M^{-1}\big)_{jl} & \big(M^{-1}\big)_{jk} \\
\end{array}
\right]
\,\phi^{(2)*}_i(x,\sigma)\,\phi^{(2)*}_j(x',\sigma')\,
\phi^{(1)}_k(x'',\sigma'')\, \phi^{(1)}_l(x''',\sigma''')
\end{multline}
In a practical simulation, the averages are performed by means of Monte
Carlo techniques. A description of the details of the numerical
algorithm used is given in Appendix \ref{app:MC}.

\section{Monte Carlo results for the correlation functions}
 \label{sec:results}

A Monte Carlo code based on the stochastic approach described in the previous section has been used
to numerically compute the expectation values of some one- and
two-body correlation functions for a one-dimensional Fermi gas with
attractive binary interactions as described by the Hamiltonian
\eq{Hamilt} with $g_0<0$. The results of analogous calculations
performed with a very similar Monte Carlo algorithm have been reported
recently in~\cite{Chomaz2}.
Other Quantum Monte Carlo schemes have also been applied to the
numerical study of the fermionic Hubbard model with attractive
interactions at finite temperature.
In particular, the determinantal QMC algorithm~\cite{FermionMC} has been
used to study the correlation functions in 2D~\cite{SpinGap} and the transition temperature
to a pair condensate state in 2D~\cite{T_cMC1,T_cMC2} and in 3D~\cite{T_c3d}.

\begin{figure}[htb]
\includegraphics[width=10cm,clip]{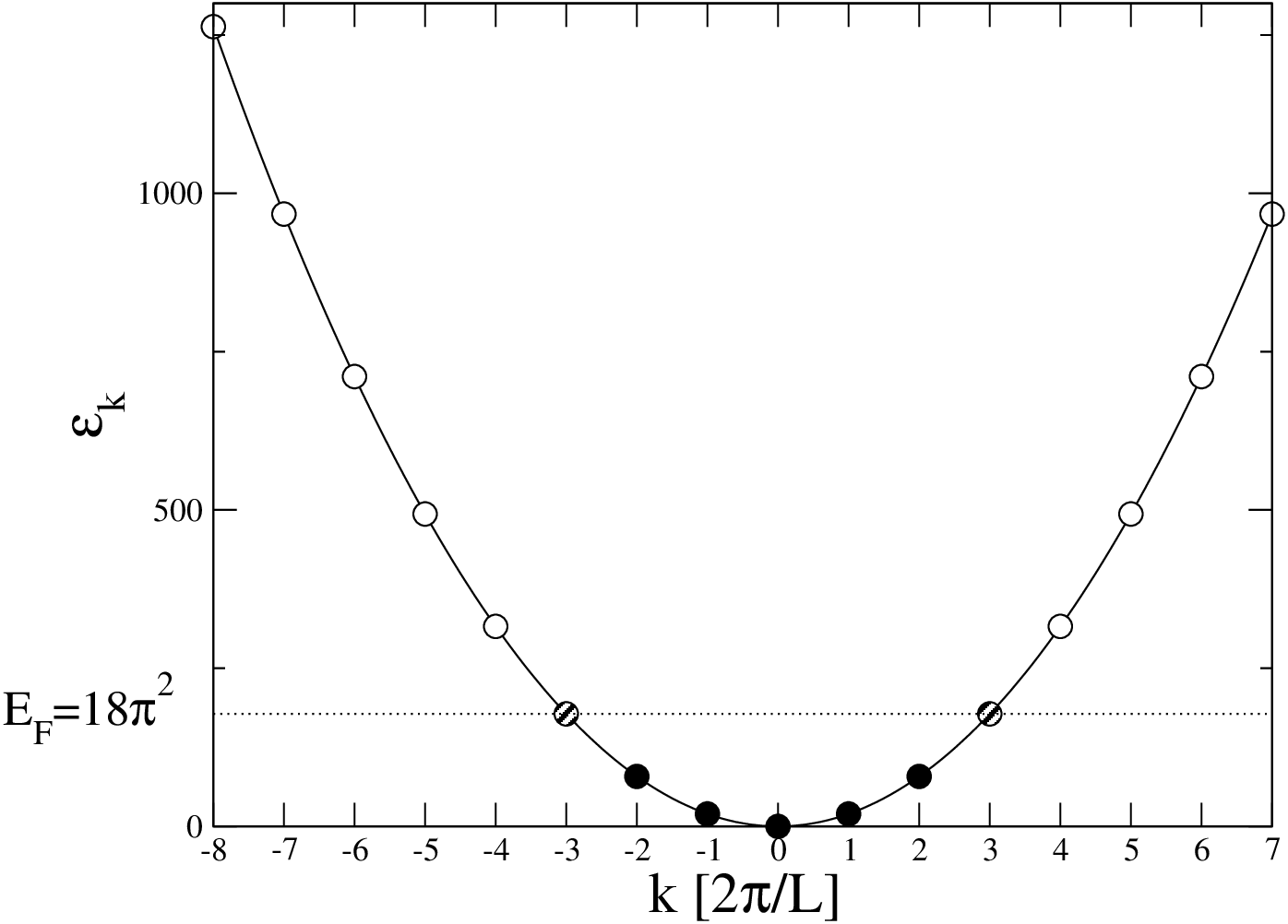}
\caption{Schematic view of the state of the ideal Fermi gas at zero
temperature for the model considered in the Monte Carlo simulation, that
is with $N=12$ atoms and ${\mathcal N}=16$ grid points. Each mode
is a plane wave with a wavevector $k=2\pi s/L$ and an energy $\epsilon_k=\hbar^2k^2/2m$, 
where the integer $s$
ranges from $-8$ to $7$. The modes with $|s|\leq 2$ are totally filled,
whereas the modes with $|s|=3$ are half-filled, the higher energy modes
being empty. $E_F=\hbar^2k_F^2/2m$ is the Fermi energy. The energies
are here in units of $\hbar^2/mL^2$. \label{fig:schema}}
\end{figure}

For our simulations, a lattice of ${\mathcal N}=16$ points was
taken, with a total number of $N=12$ atoms. A number ${\mathcal M}$ of
imaginary-time steps comprised between $400$ and $1000$ has been used.
As already mentioned, the
ensemble in which observables are calculated is the canonical one; note
that the number of particles in each of the spin state can fluctuate, only
the total number of particles is fixed. As the two spin components are
equivalent, the mean densities in each of the spin components are
equal:
\begin{equation}
\rho_\uparrow=\rho_\downarrow=\frac{N}{2L}.
\end{equation}
The state of the gas in the absence of interactions and at $T=0$
is depicted in Fig.~\ref{fig:schema}: in a given spin component,
the 5 lowest-lying single particle energy levels are totally filled,
whereas the two degenerate states of wavevectors $k_+=k_F=6\pi/L$
and $k_-=-k_F$ are half-filled. More precisely, 10 atoms are frozen
in the states of $|k|<k_F$, and the two remaining atoms are distributed
among the 4 degenerate states, $|\uparrow\,\mbox{or}\,\downarrow,\pm k_F\rangle$,
which can be done in $6$ different ways.
In presence of attractive interactions, this degeneracy
will obviously be lifted and the configurations with one atom $\uparrow$ 
and one atom $\downarrow$ with opposite momenta in the degenerate multiplicity are
favorable to the formation of a Cooper pair.

\subsection{One body correlation functions}
\begin{figure}[htbp]
\begin{center}
\includegraphics[width=10cm,clip]{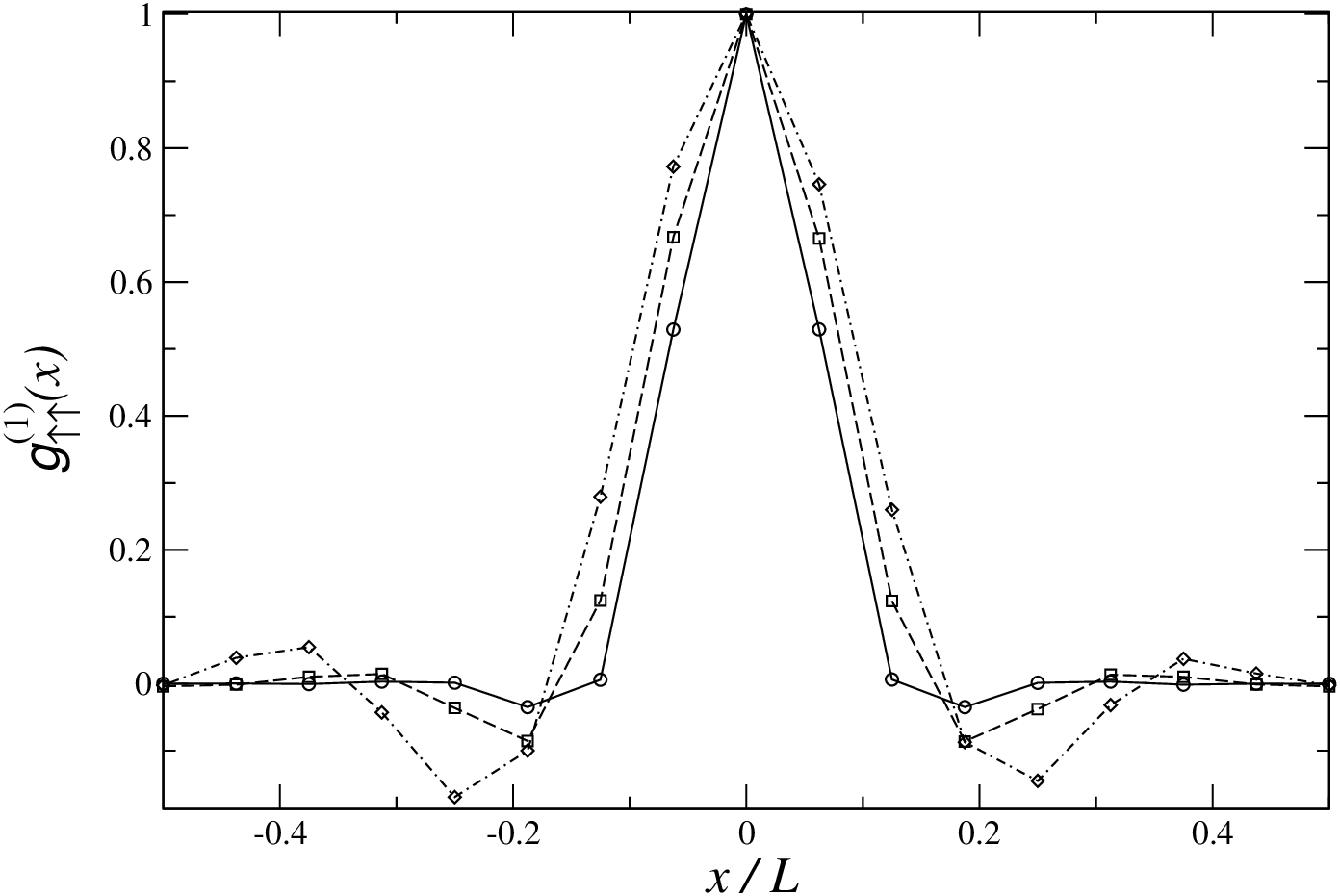}
\caption{Single-spin one-body correlation function
  $g\al{1}_{\uparrow\uparrow}(x)$ 
  for different values of the 
  temperature $T/T_F=1.12, 0.56, 0.056$ (circles, squares, diamonds). $N=12$ atoms on a
  ${\mathcal N}=16$ points lattice. Coupling constant $\rho_\uparrow
  g_0/k_B T_F=-0.42$.
\label{fig:g1uuMC}}
\end{center}
\end{figure}

The simplest observable to compute is the one-body correlation
function in a single spin state $\sigma$ (normalized to the density
$\rho_\sigma$): 
\begin{equation}
g\al{1}_{\sigma\sigma}(x)=\frac{1}{\rho_\sigma}
\big\langle\Psihd_\sigma(x)\Psih_\sigma(0) \big\rangle.
\end{equation}
The Monte Carlo prediction is plotted in fig.\ref{fig:g1uuMC} for
different values of the temperature: as expected, this correlation
function is short-ranged, coherence extending only on a length of the
order of the Fermi length $\ell_F=1/k_F$ for $T< T_F$ . 
This correlation function is indeed the Fourier
transform of the momentum distribution of the gas. 
As the interactions affect the momentum distribution only in a thin
region around the Fermi surface (the Fermi points in our
one-dimensional geometry), they do not significantly modify its
shape as compared to the ideal Fermi distribution.

Because of the rotational symmetry of the density operator in the spin space, the one-body
correlation function in different spin states:
\begin{equation}
g\al{1}_{\uparrow\downarrow}(x)=\frac{1}{\sqrt{\rho_\upa \rho_\doa}}
\big\langle\Psihd_\uparrow(x)\Psih_\downarrow(0)
\big\rangle
\end{equation}
is instead always identically vanishing.

\subsection{Density-density correlation functions}

\begin{figure}[htbp]
\begin{center}
\includegraphics[width=8cm,clip]{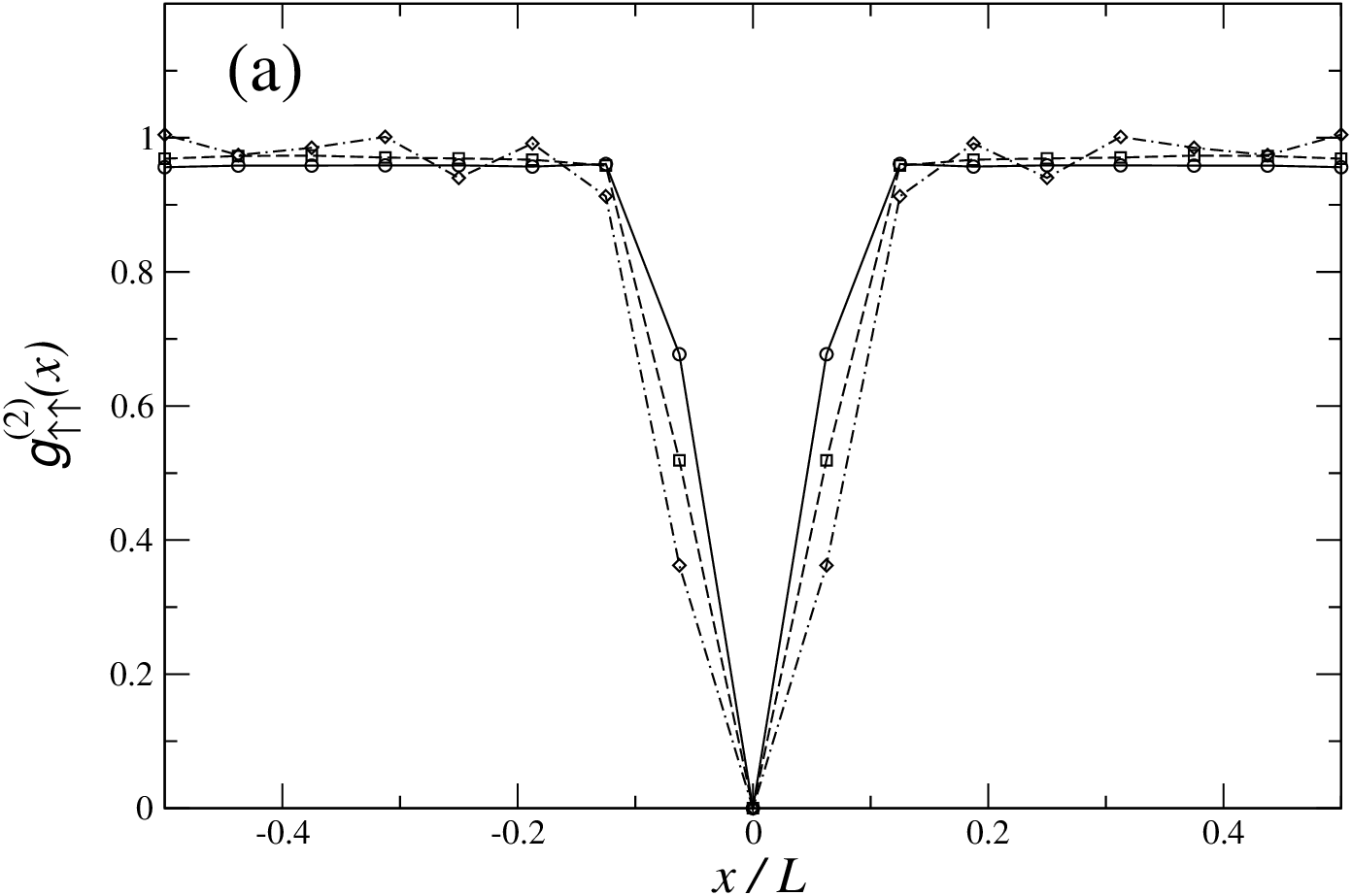}
\hspace{1cm}
\includegraphics[width=8cm,clip]{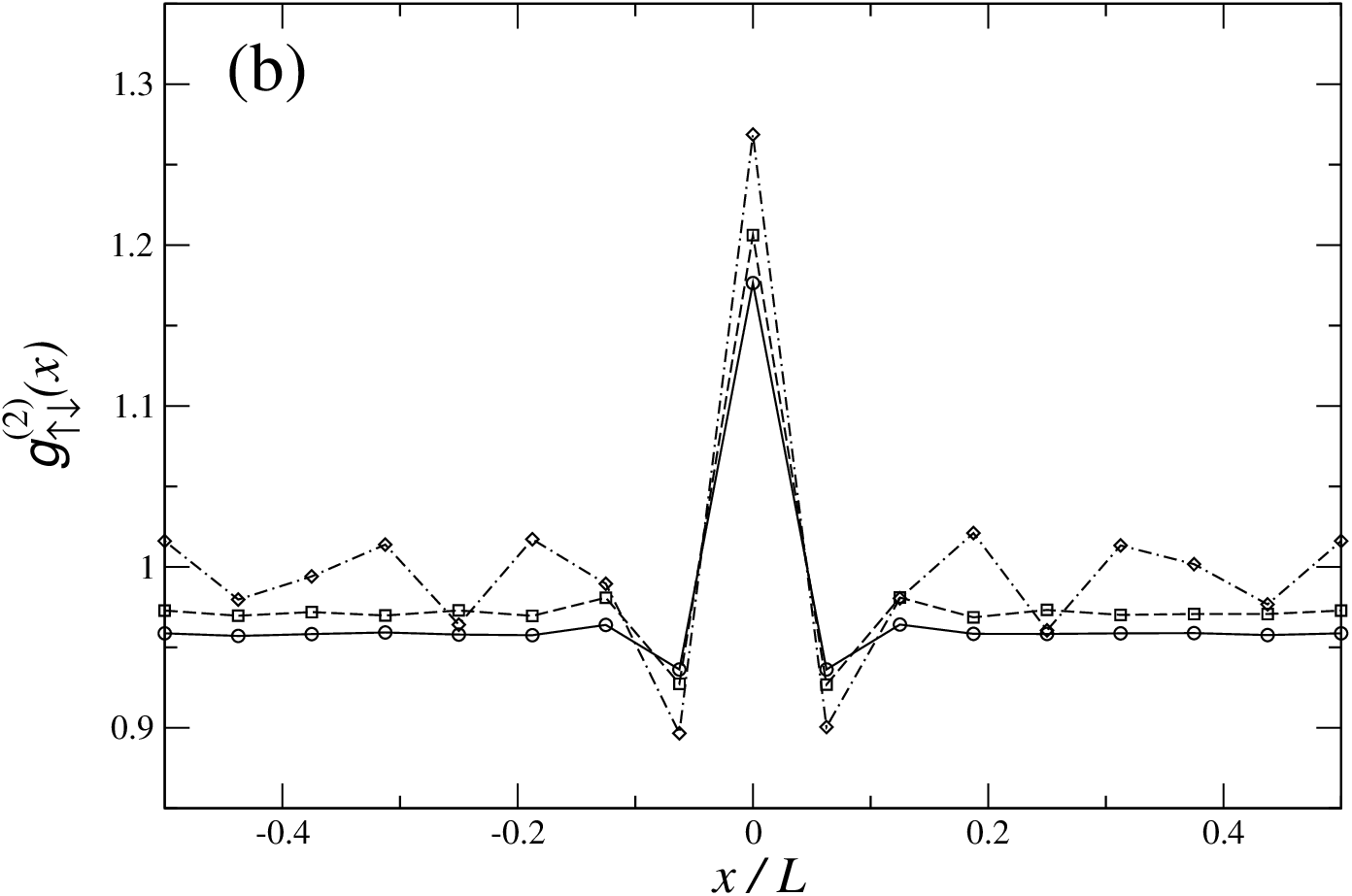}
\caption{Single-spin (a) and opposite-spin (b)
density-density correlation function
  $g\al{2}_{\uparrow\uparrow}(x)$ and $g\al{2}_{\uparrow\downarrow}(x)$
  for different values of the 
  temperature $T/T_F=1.12, 0.56, 0.056$ (circles, squares, diamonds). 
Same system parameters as in fig.\ref{fig:g1uuMC}.
\label{fig:g2MC}}
\end{center}
\end{figure}

\begin{figure}[htbp]
\begin{center}
\includegraphics[width=10cm,clip]{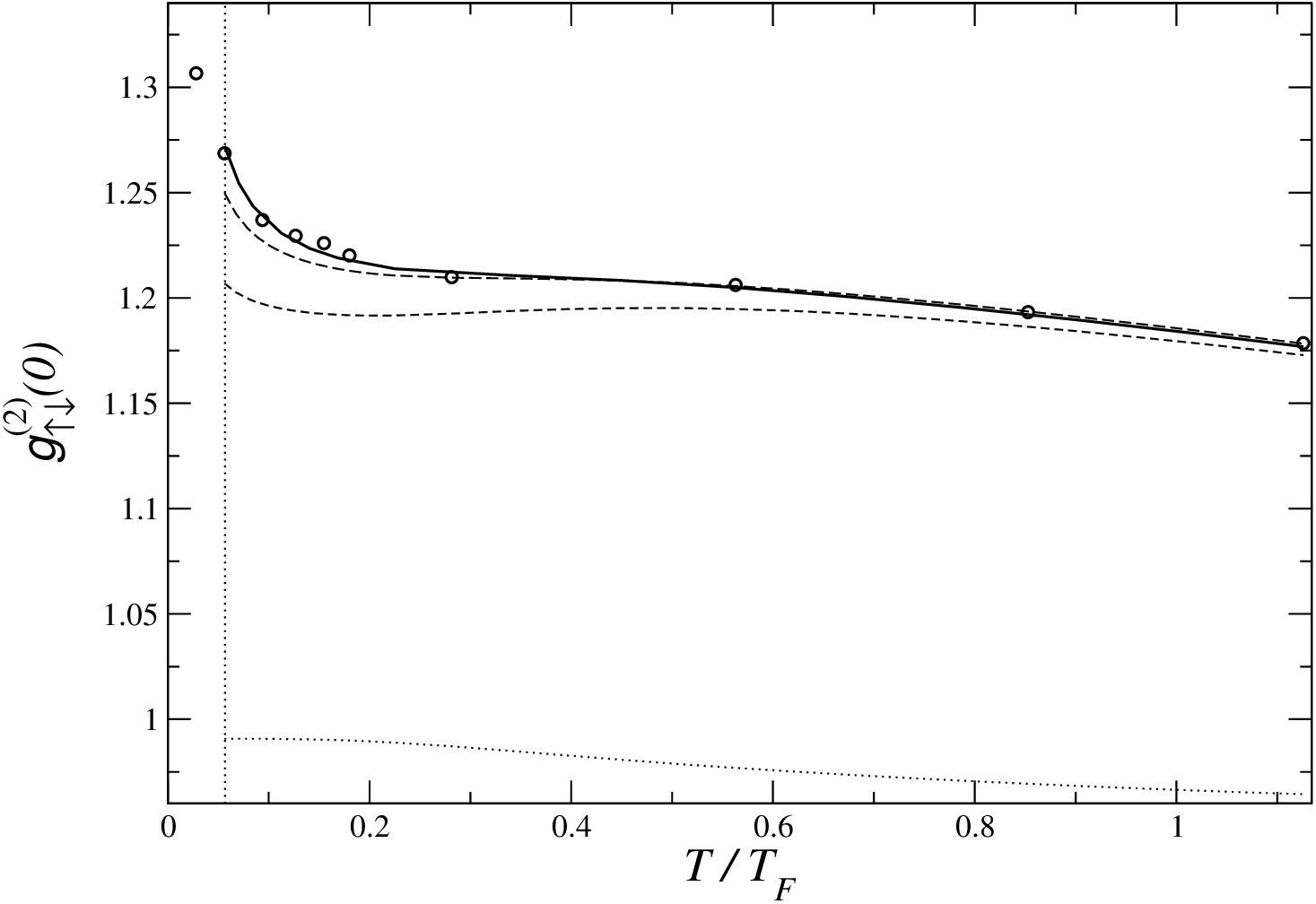}
\caption{Opposite spin density-density correlation function
  at $x=0$: $g\al{2}_{\updown}(0)$
  as a function of the temperature $T$ in the canonical ensemble. 
 Circles: Monte Carlo results. Dotted, short dashed, long dashed, solid lines:
  perturbative results upto order respectively 0, 1, 2, and 3. Same
  system parameters as in fig.\ref{fig:g1uuMC}. The vertical line is the somewhat
  arbitrary lower bound on the temperature range where perturbation theory 
  converges rapidly.
\label{fig:g2udC_T}}
\end{center}
\end{figure}

Density-density correlation functions are another observable of
interest. Both the single-spin density-density correlation function:
\begin{equation}
g\al{2}_{\sigma\sigma}(x)=\frac{1}{\rho_\sigma^2}\big\langle
\Psihd_\sigma(0)\Psihd_\sigma(x) 
\Psih_\sigma(x)\Psih_\sigma(0) \big\rangle =
\frac{1}{\rho_\sigma^2} \big\langle \hat{\rho}_\sigma(x)\hat{\rho}_\sigma(0)\big\rangle 
-\frac{1}{\rho_\sigma\,dx}\,\delta_{x,0}
\end{equation}
and the opposite-spin one:
\begin{equation}
g\al{2}_{\uparrow\downarrow}(x)=\frac{1}{\rho_\upa \rho_\doa}
\big\langle\Psihd_\uparrow(0)\Psihd_\downarrow(x)
\Psih_\downarrow(x)\Psih_\uparrow(0) \big\rangle
=\frac{1}{\rho_\uparrow \rho_\downarrow} \big\langle \hat{\rho}_\uparrow(x) \hat{\rho}_\downarrow(0)\big\rangle,
\end{equation}
with $\hat{\rho}_\sigma(x)\equiv \hat{\Psi}^\dagger_\sigma(x)\hat{\Psi}_\sigma(x)$,
have been calculated by Monte Carlo and plotted as a function of $x$ respectively in
fig.\ref{fig:g2MC}a and in fig.\ref{fig:g2MC}b.
In fig.\ref{fig:g2udC_T}, we have plotted
$g\al{2}_{\uparrow\downarrow}(0)$ as a function of temperature. 
The magnitude of actual density correlations is quantified by the
difference $g\al{2}_{\sigma\sigma'}(x)-1$. 

On one hand, the density correlations in a single spin state described
by $g\al{2}_{\uparrow\uparrow}$ show a short-range hole (Pauli hole)
of width similar to the bump of the one-body correlation function
$g\al{1}_{\uparrow\uparrow}$ and are weakly affected by the
interactions and by the temperature variations (fig.\ref{fig:g2MC}a). 

On the other hand, the density  correlations between opposite spins
described by $g\al{2}_{\uparrow\downarrow}$ show an interesting
temperature dependence in the presence of attractive interactions. 
The lower is the temperature, the most effective are in fact the
interactions and therefore the stronger the bunching of opposite spin
particles on a given lattice site.
In fig.\ref{fig:g2MC}b we have plotted the spatial profile of
$g\al{2}_{\upa\doa}(x)$ for different values of the temperature: for
the lowest value of $T/T_F$, notice not only the increase of
$g\al{2}_{\upa\doa}(0)$, but also the appearance of oscillations as a
function of $x$.
As we shall see in the next subsection, at this temperature a
condensate of pairs is present.
The oscillations then result from the contribution of two distinct
effects: the 
Friedel oscillations in the correlation functions of the
normal phase which follow from the sharpness of the Fermi
surface~\cite{Mahan}, and the oscillations shown by the Cooper pair
wavefunction described within the BCS theory by the pairing function 
$\langle\hat{\Psi}_\uparrow(x)\hat{\Psi}_\downarrow(0)\rangle$.
In fig.\ref{fig:g2udC_T} we have summarized the values of
$g\al{2}_{\uparrow\downarrow}(0)$ as a function of the temperature.  
Notice that $g\al{2}_{\upa\doa}(0)-1$ is appreciable already at the
highest temperature considered in fig.\ref{fig:g2udC_T}, which, as we
shall see in the next subsection, is much higher than the critical
temperature $T^*$ for the appearance of long-range order.

\subsection{First-order pair coherence function}

It is believed in statistical physics that the superfluid
transition in two-component Fermi systems with attractive binary
interactions is related to the appearance of long-range order 
in the so-called {\it anomalous} averages~\cite{LandauCM}.
In symmetry breaking theories such as the BCS one, this feature
corresponds to a non-vanishing value for the gap function defined as: 
\begin{equation}
\Delta=-g_0 \langle \Psih_\upa(x)\Psih_\doa(x) \rangle,
\eqname{Anomalous}
\end{equation}
which plays the role of the order parameter of the phase transition in
a Ginzburg-Landau approach.
In number conserving approaches, quantities like \eq{Anomalous} are zero.
The phase transition however still appears in the long-range behaviour
of correlation functions of the form: 
\begin{equation}
g\al{1}_{\rm pair}(x)=\frac{1}{\rho_\upa \rho_\doa}
\big\langle
\Psihd_\doa(x)\,\Psihd_\upa(x)\,
\Psih_\upa(0)\,\Psih_\doa(0) \big\rangle.
\eqname{LR_Anomalous}
\end{equation}
A similar criterion was used in~\cite{T_cMC1,T_cMC2,T_c3d} to determine the
transition temperature. 

A simple physical interpretation of $g\al{1}_{\rm pair}$ can be
provided as the first order correlation function of pairs: the
operator $\Psih_\upa(0)\Psih_\doa(0)$ annihilates in fact a pair of
particles in opposite spin states at the spatial position $0$ and
the operator $\Psihd_\doa(x)\Psihd_\upa(x)$ creates them
back at $x$. This correlation function is therefore formally equivalent to the
first order coherence function of a composite boson formed by a pair of fermions
with opposite spins. From this point of view, the non-vanishing long-range
limit of $g\al{1}_{\rm pair}(x)$ is a signature of a quantum condensation of pairs.

\begin{figure}[htbp]
\begin{center}
\includegraphics[width=8cm,clip]{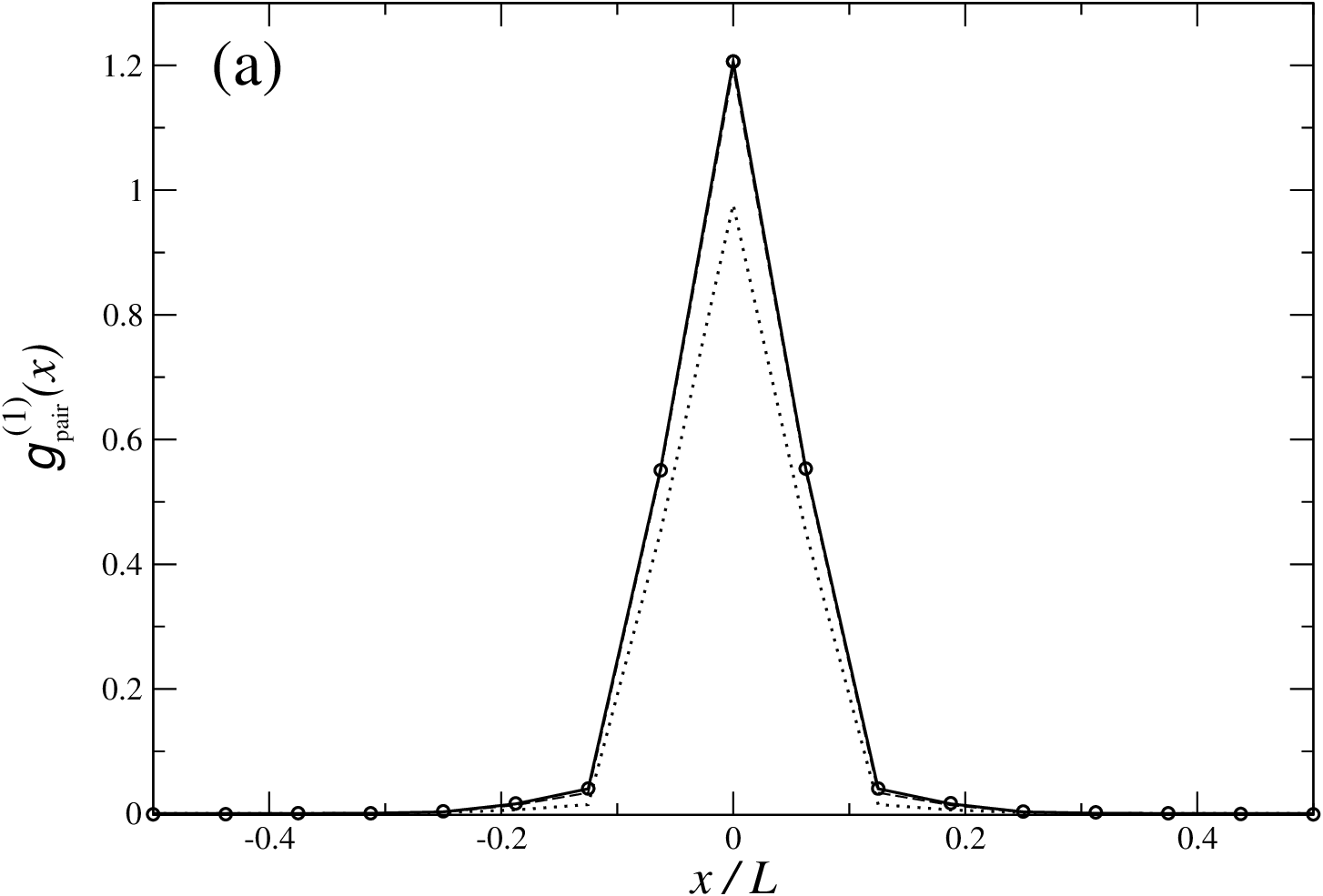}
\hspace{1cm}
\includegraphics[width=8cm,clip]{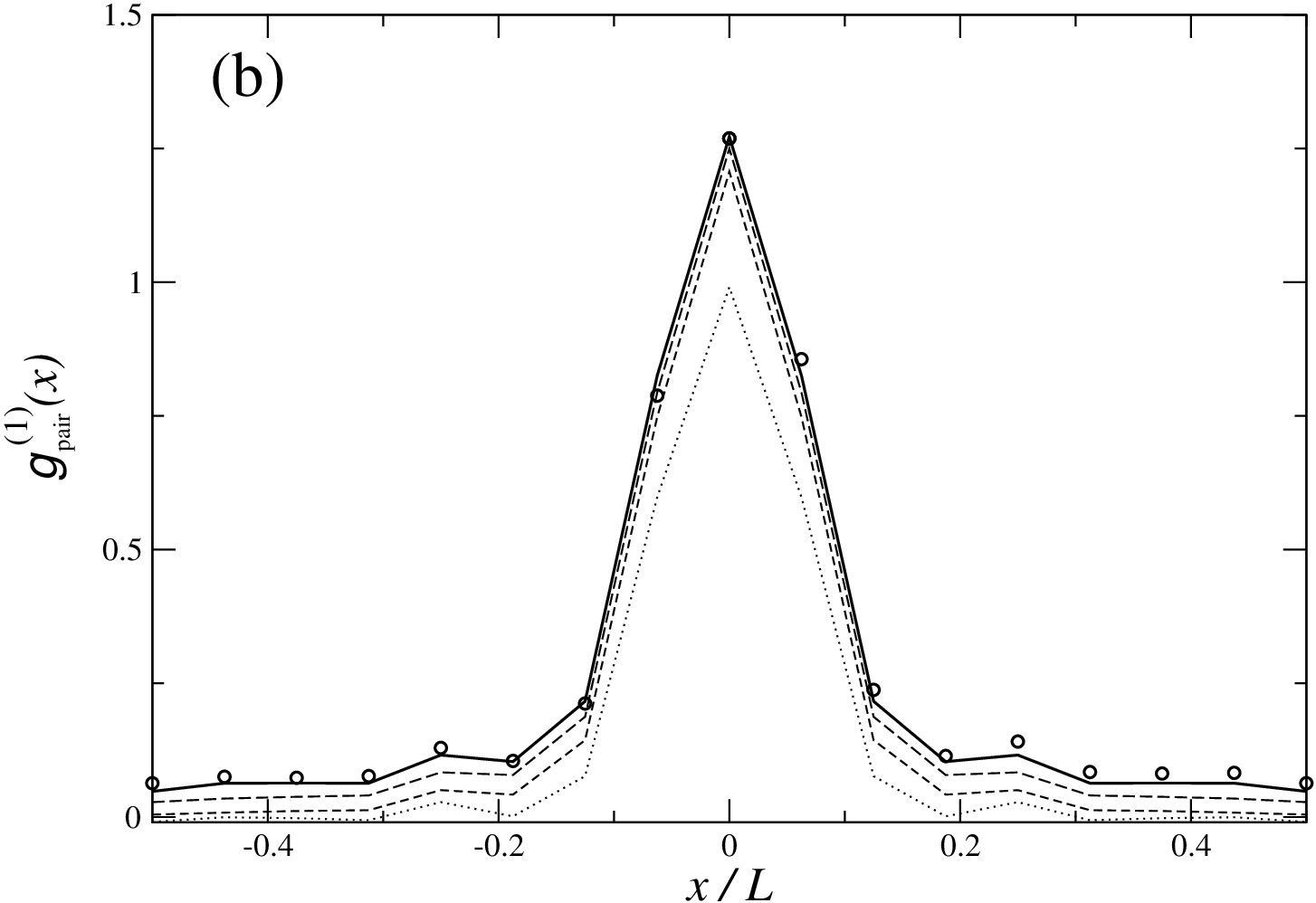}
\caption{
Normalized pair coherence function
  $g\al{1}_{\rm pair}(x)$
  for two different temperatures $T=0.56 T_F$ (left panel) and $T=0.056 T_F$
  (right panel) in the canonical ensemble. 
Circles: Monte Carlo results. Dotted,
  dashed, solid lines in left panel: perturbative results upto order
  respectively 0, 1 and 2 (orders 1 and 2 are
  undistinguishable). Dotted, short dashed, long dashed, solid lines
  in right panel: perturbative results upto order
  respectively 0, 1, 2, and 3. Same
  system parameters as in fig.\ref{fig:g1uuMC}.
\label{fig:g1pairMC}}
\end{center}
\end{figure}

\begin{figure}[htbp]
\begin{center}
\includegraphics[width=10cm,clip]{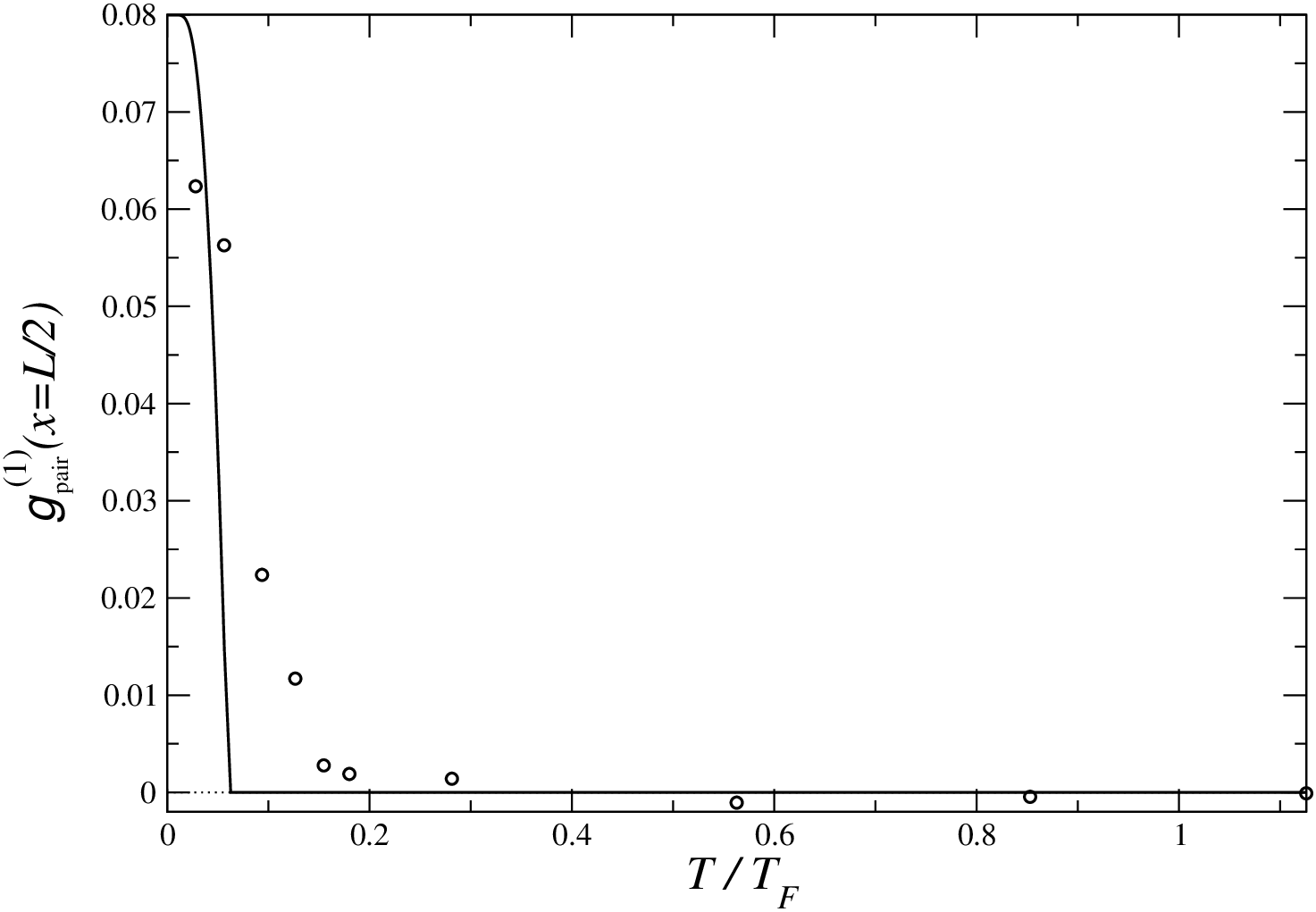}
\caption{
\label{fig:g1pairC_T}
Pair coherence function
  $g\al{1}_{\rm pair}(L/2)$
  as a function of the temperature $T$.
Circles: Monte Carlo results in the canonical ensemble.  Solid line:
  BCS theory. 
Same  system parameters as in fig.\ref{fig:g1uuMC}.}
\end{center}
\end{figure}

Monte Carlo simulations for this quantity are shown in
fig.\ref{fig:g1pairMC}. At low temperatures, 
$g\al{1}_{\rm pair}(x)$ has a finite value also for $x=L/2$, i.e. at
the largest distance from $0$ allowed by the finite size of the
box. On the other hand, at higher temperatures, but still much lower
than the Fermi temperature, $g\al{1}_{\rm pair}(L/2)$ becomes very
small and the long-range order is destroyed.
To make this cross-over more apparent, we have plotted in Fig.~\ref{fig:g1pairC_T}
the value of $g\al{1}_{\rm pair}(L/2)$ as a function of the temperature: a sudden
rise of this quantity appears at low temperatures.
This behavior qualitatively corresponds to the one expected for a BCS
transition: although a BCS transition can not occur in one
dimension in the thermodynamical limit because of long wavelength fluctuations
destroying the long range order \cite{1Dg1pair}, it can however be observed in our
simulations because of the finite size of the system. As the system is
finite, the transition temperature $T^*$ is not precisely defined and the
long-range order has an analytic dependance on temperature.
Notice that the opposite spin density-density correlation described by
$g\al{2}_{\uparrow\downarrow}(0)$ are already important at
$T>T^*$ and for $T<T^*$ they only get slightly reinforced. 

\subsection{Second-order momentum space correlation function}

Another observable that has been recently proposed as a possible way
of detecting the transition to a pair condensate state is the
second-order momentum space correlation function~\cite{Lukin}:
\begin{equation}
\eqname{Lukin_Obs}
G\al{2}_{k}(k)=\big\langle {\hat n}_{k\upa}\,{\hat n}_{-k\doa}\big\rangle-
\big\langle {\hat n}_{k\upa} \big\rangle \big\langle {\hat n}_{-k\doa}\big\rangle,
\end{equation}
where the operator ${\hat n}_{k\sigma}=\ahd_{k\sigma}\ah_{k\sigma}$ gives the occupation of the
plane wave $k$ with spin component $\sigma$.

As discussed in~\cite{Lukin}, BCS theory predicts that correlations should be
absent above $T_{\mathrm{BCS}}$, that is $G\al{2}_k=0$, while the transition to a
condensate state  should be observable as the appearance of a
non-vanishing value of $G\al{2}_k$, sharply peaked around $k=k_F$.
\begin{figure}[htbp]
\begin{center}
\includegraphics[width=10cm,clip]{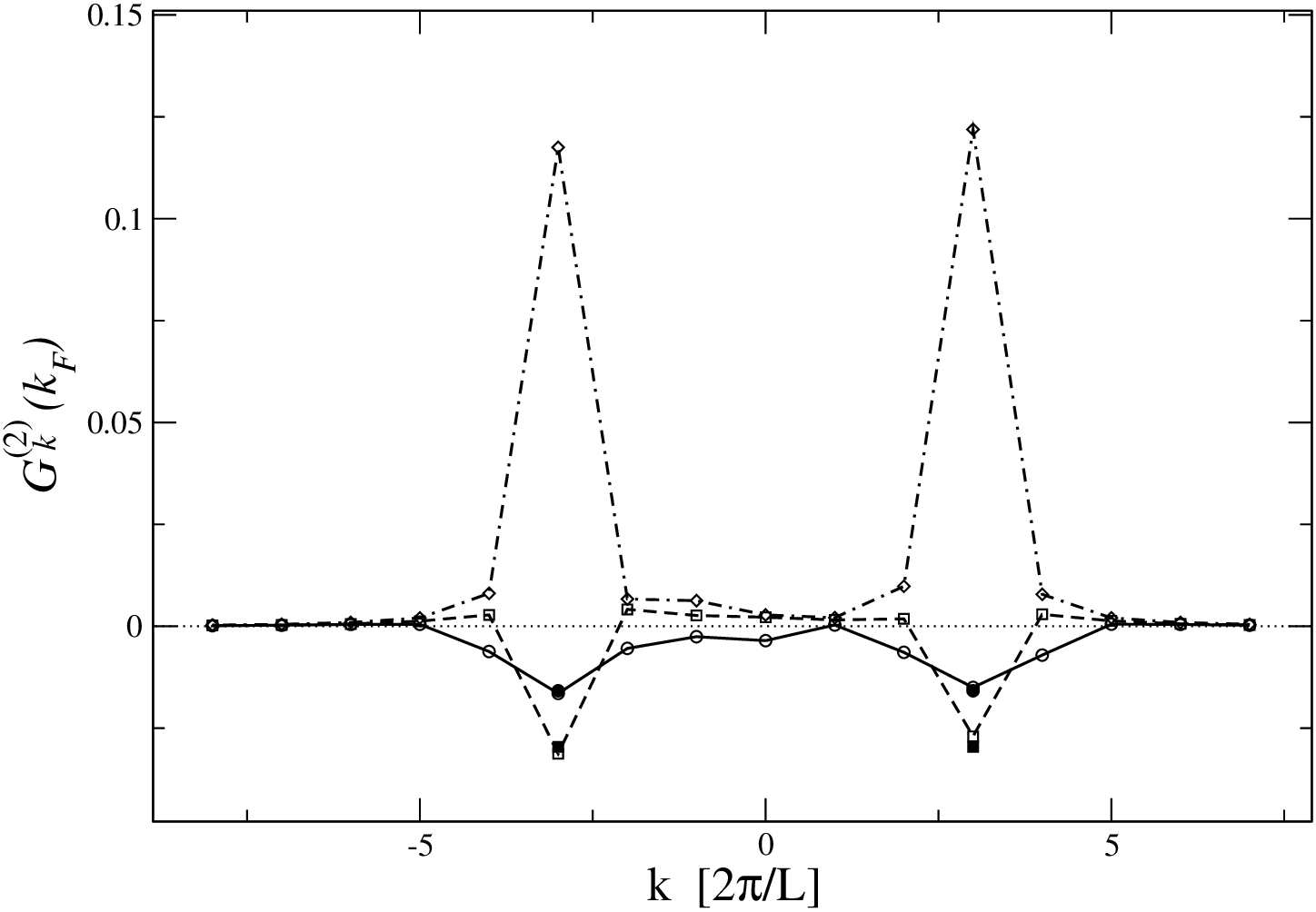}
\caption{
\label{fig:g2kkMC}
Second-order momentum space correlation function $G\al{2}_k(k)$ 
  for different values of the 
  temperature.
Empty circles, squares, diamonds: Monte Carlo results for $T/T_F=0.56, 0.28, 0.056$.
Filled circles and squares at $k=\pm k_F$: perturbative expansion up
  to order 3 for
  $T/T_F=0.56, 0.28$ respectively. For $T/T_F=0.056$ a perturbative
  expansion to an order higher than 3 would be required to observe convergence.
}
\end{center}
\end{figure}

\begin{figure}[htbp]
\begin{center}
\includegraphics[width=10cm,clip]{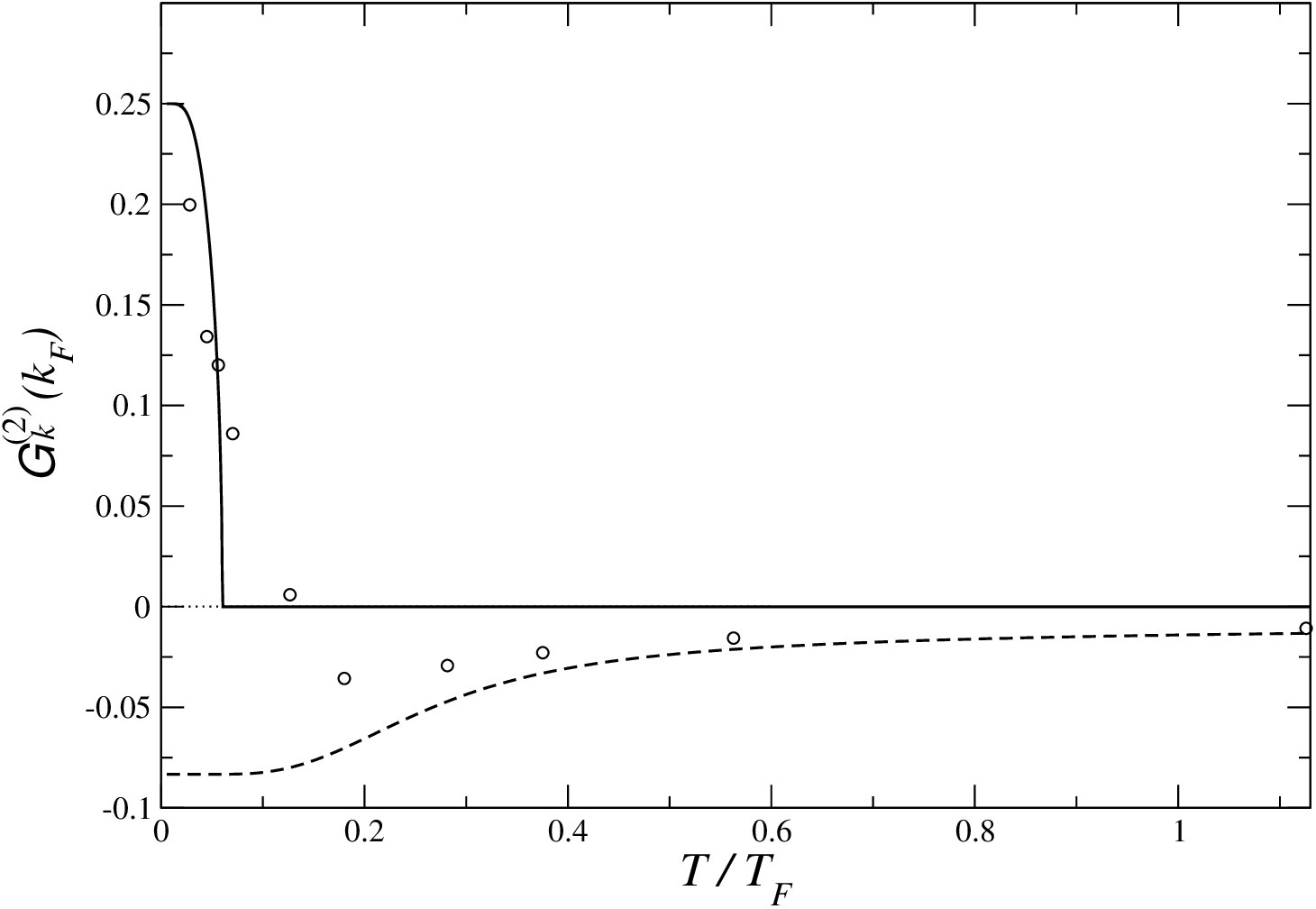}
\caption{
\label{fig:g2kkC_T}
Second-order momentum space correlation function at $k=k_F$: $G\al{2}_k(k_F)$
as a function of the temperature $T$. 
Circles: Monte Carlo results in the canonical ensemble.  Solid line:
BCS theory in the grand canonical ensemble. Dashed line: ideal gas in
the canonical ensemble.
Same  system parameters as in fig.\ref{fig:g1uuMC}.}
\end{center}
\end{figure}

In fig.\ref{fig:g2kkMC}, we have plotted Monte Carlo results for
$G\al{2}_k$ as a function of $k$ for different values of the
temperature. At all temperatures, the quantity is indeed strongly peaked at
$k=k_F$, and nearly vanishes at the other values.
A summary of the temperature-dependence of $G\al{2}_k(k_F)$ is plotted
in fig.\ref{fig:g2kkC_T}.
At temperatures above the transition temperature $T^*$, correlations
are negative and increase as the temperature is lowered. The negative
correlation simply follows from the fact that we are working in the
canonical ensemble, that is at a fixed total number of particles (see
the ideal Fermi gas result in fig.\ref{fig:g2kkC_T}).
As the temperature drops below $T^*$, the correlations change sign
becoming large and positive.
This is a signature of pairing: because of the attractive
interactions, the states with a filled Fermi sphere  
plus two particles in states of opposite momenta and spins are
in fact energetically favoured. In this state, the fluctuations of the
occupation numbers of the $k_F,\upa$ and $-k_F,\doa$ states are
positively correlated.

\section{Results of a perturbative expansion in $g_0$}
\label{sec:Pert}

In this section, we explain how to calculate the pair coherence function $g_{\rm pair}^{(1)}$ and
the density correlation function $g^{(2)}_{\uparrow\downarrow}$ by means of a
series expansion in powers of the  coupling constant $g_0$.
The same procedure was used in fig.\ref{fig:g2kkMC} to obtain a series
expansion for the momentum-space second order correlation function $G\al{2}_k$
although we do not give here the details of the calculation.
We expect this perturbative approach performed around the ideal Fermi gas to be efficient
mainly at $T>T^*$ that is in absence of a condensate of pairs. For $T<T^*$ we indeed found numerically
that the series (up to order 3) is slowly convergent. Note that such a series expansion can however
be shown to be convergent at non-zero temperature for our model system with a finite number of modes,
see below.

\subsection{In the canonical ensemble}

As we wish to compare to the Quantum Monte Carlo results, we have in principle
to perform the perturbative treatment directly in the canonical ensemble with
$N$ particles. The resulting averages in the ideal Fermi gas thermal
state are however difficult 
to evaluate analytically. We therefore apply the following trick to `canonize' the
grand canonical ensemble. We introduce an unnormalized grand
canonical thermal density operator defined as
\begin{equation}
\sigma_{\rm gc}(\theta) = e^{-\beta({\mathcal H}-\mu_0 \hat{N})} e^{i\theta\hat{N}}
\end{equation}
where $\hat{N}$ is the total number operator, $\theta$ is an angle and
$\mu_0$ is the chemical potential of the ideal Fermi gas with an average number of $N$ 
particles. Taking the Fourier component of $\sigma(\theta)$ over the
harmonic $e^{i\theta N}$ 
amounts to projecting $\sigma(\theta)$ over the subspace with exactly $N$ particles.
The canonical expectation value of an operator $O$ is therefore
exactly given by
\footnote{In practice, a numerically more efficient formulation
can be obtained from the fact that the total number of spin 1/2
fermions for a given spatial grid with ${\mathcal N}$  points has an
upper limit of $2{\mathcal N}$.  If one excludes the cases $N=0$ and
$N=2{\mathcal N}$, one can replace the integrals over $\theta$ by
discrete sums over the values $\theta = 2\pi q/(2{\mathcal N})$ where
the integer $q$ ranges from $0$ to $2{\mathcal N}-1$. Furthermore, the
symmetry $\sigma_{\rm gc}(\theta)^\dagger=\sigma_{\rm gc}(-\theta)$
may be used to reduce the range of $q$.}:
\begin{equation}
\langle O\rangle_N = \frac{\int_0^{2\pi} d\theta\, e^{-i\theta N}
  \mbox{Tr}[\sigma_{gc}(\theta) O]} 
{\int_0^{2\pi} d\theta\, e^{-i\theta N} \mbox{Tr}[\sigma_{gc}(\theta)]}.
\label{eq:can_afo_gc}
\end{equation}

We then expand $e^{-\beta {\mathcal H}}$ in powers of the interaction
potential $V$, here up to third order:
\begin{eqnarray}
e^{-\beta ({\mathcal H}-\mu_0 \hat{N})} &=& e^{-\beta ({\mathcal H}_0-\mu_0\hat{N})} 
\left[1 -\int_0^{\beta}d\tau_1\, V(\tau_1)
  +\int_0^{\beta}d\tau_2\int_0^{\tau_2}d\tau_1\, V(\tau_2) V(\tau_1)
  \right.\nonumber\\ 
&&\left. -
  \int_0^{\beta}d\tau_3\int_0^{\tau_3}d\tau_2\int_0^{\tau_2}d\tau_1\,
  V(\tau_3) V(\tau_2) V(\tau_1) 
+\ldots
\right]
\label{eq:expansion}
\end{eqnarray}
where ${\mathcal H}_0$ is the kinetic energy operator of the gas and
the imaginary time interaction picture 
for an operator $X$ is defined as
\begin{equation}
X(\tau) = e^{\tau ({\mathcal H}_0-\mu_0\hat{N})} X e^{-\tau ({\mathcal
    H}_0-\mu_0\hat{N})}. 
\end{equation}
For a non-zero temperature and a finite number of grid points,
the norm of the operator $V(\tau)$ is finite as both $V$ and ${\mathcal
H}_0-\mu_0\hat{N}$ have a finite norm.
As a consequence, the norm of the $n^{\rm th}$-order contribution to the series
expansion Eq.~(\ref{eq:expansion}) 
can be bounded from above by $A^n/n!$ where $A$ is some number, and
the series Eq.~(\ref{eq:expansion}) 
is absolutely convergent 
\footnote{One can take e.g. $A=||V||(e^{\beta\Delta
E_0}-1)/\Delta E_0$ where $\Delta E_0$ is the difference between the
largest and the smallest eigenvalues of ${\cal H}_0-\mu_0\hat{N}$ and
where the norm of $V$ is $||V||\leq |g_0|L/dx^2$.}.

The calculation of the numerator of Eq.~(\ref{eq:can_afo_gc}) then
involves the $\theta$-dependent grand canonical partition 
function of the ideal Fermi gas and $\theta$-dependent  expectation
values in the grand canonical ideal Fermi gas:
\begin{eqnarray}
\Xi_0(\theta) &\equiv& \mbox{Tr}[\sigma_{\rm gc}^{0}(\theta)] =
\prod_{k}
\left(1+e^{-\beta[\hbar^2k^2/(2m)-\mu_0]}e^{i\theta}\right)^2 \\ 
\langle X\rangle_0(\theta) &\equiv&
\frac{1}{\Xi_0(\theta)}\mbox{Tr}[\sigma_{\rm gc}^0(\theta)X] 
\end{eqnarray}
where the square originates from the presence of two spin components.
The operator $X$ is one of the terms inside the square brackets of
Eq.\eq{expansion}. 
The expectation values can be evaluated using Wick's theorem and
involve the following particle and hole 
correlation functions:
\begin{eqnarray}
G_0(x,\tau;\theta) &=& \langle \hat{\psi}_{\uparrow}^\dagger(x,\tau)
\hat{\psi}_{\uparrow}(0)\rangle_0(\theta) \\ 
\bar{G}_0(x,\tau;\theta) &=& \langle \hat{\psi}_{\uparrow}(x,\tau)
    \hat{\psi}_{\uparrow}^\dagger(0)\rangle_0(\theta). 
\end{eqnarray}
The explicit expressions of the relevant expectation values in terms 
of $G_0$ and $\bar{G}_0$ are given in the Appendix \ref{appen:expect}.
The integrals over the `times' $\tau_1,\tau_2,\tau_3$ 
and the sums over the grid points associated to each factor
$V(\tau_i)$ are performed numerically. As each integral is discretized
in 256 steps and there are 16 grid points 
in the lattice, the calculation of the third order correction involves
the summation of about $10^{10}$ terms 
for a given value of $\theta$.

The perturbative results for the $x=0$ pair distribution function
$g^{(2)}_{\uparrow\downarrow}(0)$  
are plotted against the Monte Carlo results as
a function of temperature in Fig.~\ref{fig:g2udC_T} 
for various orders of the perturbative
expansion. The agreement with the second 
order expansion is perfect at high temperature, whereas the third
order contribution is required to have agreement 
at lower temperatures 
\footnote{Each term of the perturbative
  expansion is expected to diverge in the $T\rightarrow 0$ 
limit; this can be checked to be the case for the first order correction to 
$g^{(2)}_{\uparrow\downarrow}(0)$ 
in the grand canonical ensemble,
this correction diverging as $-\beta g_0/L$. We therefore restrict the perturbative expansion
to temperatures larger than $|g_0|/L\sim 10$ in dimensionless units.}. 
For a given temperature, the $x$ dependence of
$g^{(2)}_{\uparrow\downarrow}(x)$ predicted 
by the perturbative expansion is also in good agreement with the exact
Monte Carlo results. For $g_{\rm pair}^{(1)}(x)$, the agreement is also 
good, at high temperature in Fig.~\ref{fig:g1pairMC}a, as well as at
a temperature $T<T^*$ in Fig.~\ref{fig:g1pairMC}.
The fact that the third order prediction is very close to the Quantum Monte Carlo
results even when long range order is present may be fortuitous: it significantly differs
from the second order prediction so that a calculation of 
the fourth order correction is required to justify the truncation of the series
at this order.

\subsection{In the grand canonical ensemble}

It is actually interesting to perform also the perturbative expansion
in the grand canonical 
ensemble: simpler analytical formulas can be obtained, which can be used
to test existing approximate 
theories applicable to the grand canonical ensemble.
The unnormalized density operator of the gas is now
\begin{equation}
\sigma_{\rm gc} = e^{-\beta({\mathcal H}-\mu\hat{N})}.
\label{eq:sigma_gc}
\end{equation}
The perturbative expansion has to be performed for a fixed value of
the mean total number of 
particles equal to $N$. As a consequence the value of the chemical
potential $\mu$ is not known 
in advance and has to be adjusted order by order in the perturbative
expansion. To this end, we write 
\begin{equation}
\mu = \mu_0  + \delta \mu
\end{equation}
where $\mu_0$ is the chemical potential of the ideal Fermi gas having
on the mean a number 
$N$ of particles.  This amounts to performing the following splitting:
\begin{equation}
{\mathcal H}-\mu\,\hat{N} = ({\mathcal H}_0 -\mu_0\hat{N}) + W
 \end{equation}
where the perturbation is now
\begin{equation}
W = V -\delta\mu \, \hat{N},
\end{equation}
both terms in $W$ being of order $g_0$.
We shall restrict here for simplicity to a second order
expansion. From Eq.(\ref{eq:expansion}) 
we get
\begin{equation}
\langle O\rangle = \frac{\langle O\rangle_0 -\int_0^{\beta}d\tau_1
  \langle W(\tau_1)O\rangle_0 
+\int_0^{\beta} d\tau_2\int_0^{\tau_2} d\tau_1\, \langle
  W(\tau_2)W(\tau_1)O\rangle_0 +\ldots} 
{1-\int_0^{\beta}d\tau_1 \langle W(\tau_1)\rangle_0+
\int_0^{\beta} d\tau_2\int_0^{\tau_2} d\tau_1\, \langle
  W(\tau_2)W(\tau_1)\rangle_0 +\ldots} 
\label{eq:exp_frac}
\end{equation}
where $\langle X \rangle=\textrm{Tr}[\sigma_{\rm
    gc}X]/\textrm{Tr}[\sigma_{\rm gc}]$ stands for the expectation value in the grand canonical 
density operator of the interacting gas Eq.~(\ref{eq:sigma_gc})
and $\langle X \rangle_0$ stands for the expectation value in the grand canonical
density operator $\exp[-\beta({\mathcal H}_0-\mu_0\hat{N})]$ of the ideal Fermi gas.
Expanding the inverse of the denominator in Eq.(\ref{eq:exp_frac}) and keeping terms
up to second order, one obtains
\begin{equation}
\langle O\rangle = \langle O\rangle_0-\int_0^{\beta}d\tau_1\,  
\langle\langle W(\tau_1) O\rangle\rangle_0
+\int_0^{\beta}d\tau_2\int_0^{\tau_2}d\tau_1\, \langle\langle
W(\tau_2) W(\tau_1) O\rangle\rangle_0 
+O(g_0^3)
\label{eq:exp_joli}
\end{equation}
where we have introduced the irreducible averages of products of
operators $A$, $B$, $C$: 
\begin{eqnarray}
\langle\langle A B\rangle\rangle_0 &\equiv& \langle AB\rangle_0
-\langle A\rangle_0 \langle B\rangle_0 \\ 
\langle\langle A B C\rangle\rangle_0 &\equiv& \langle ABC\rangle_0 - \langle A\rangle_0
\langle BC\rangle_0 - \langle B\rangle_0 \langle AC\rangle_0 - \langle
C\rangle_0 \langle AB\rangle_0  
\nonumber\\
&& +2 \langle A\rangle_0 \langle B\rangle_0 \langle C\rangle_0
\end{eqnarray}
and where we used the identity
\begin{equation}
\frac{1}{2}\left(\int_0^{\beta} d\tau_1\, \langle W(\tau_1)\rangle_0\right)^2 =
\int_0^{\beta}d\tau_2\int_0^{\tau_2}d\tau_1\, \langle W(\tau_2)\rangle_0
\langle W(\tau_1)\rangle_0.
\end{equation}

\begin{figure}[htbp]
\begin{center}
\includegraphics[width=10cm,clip]{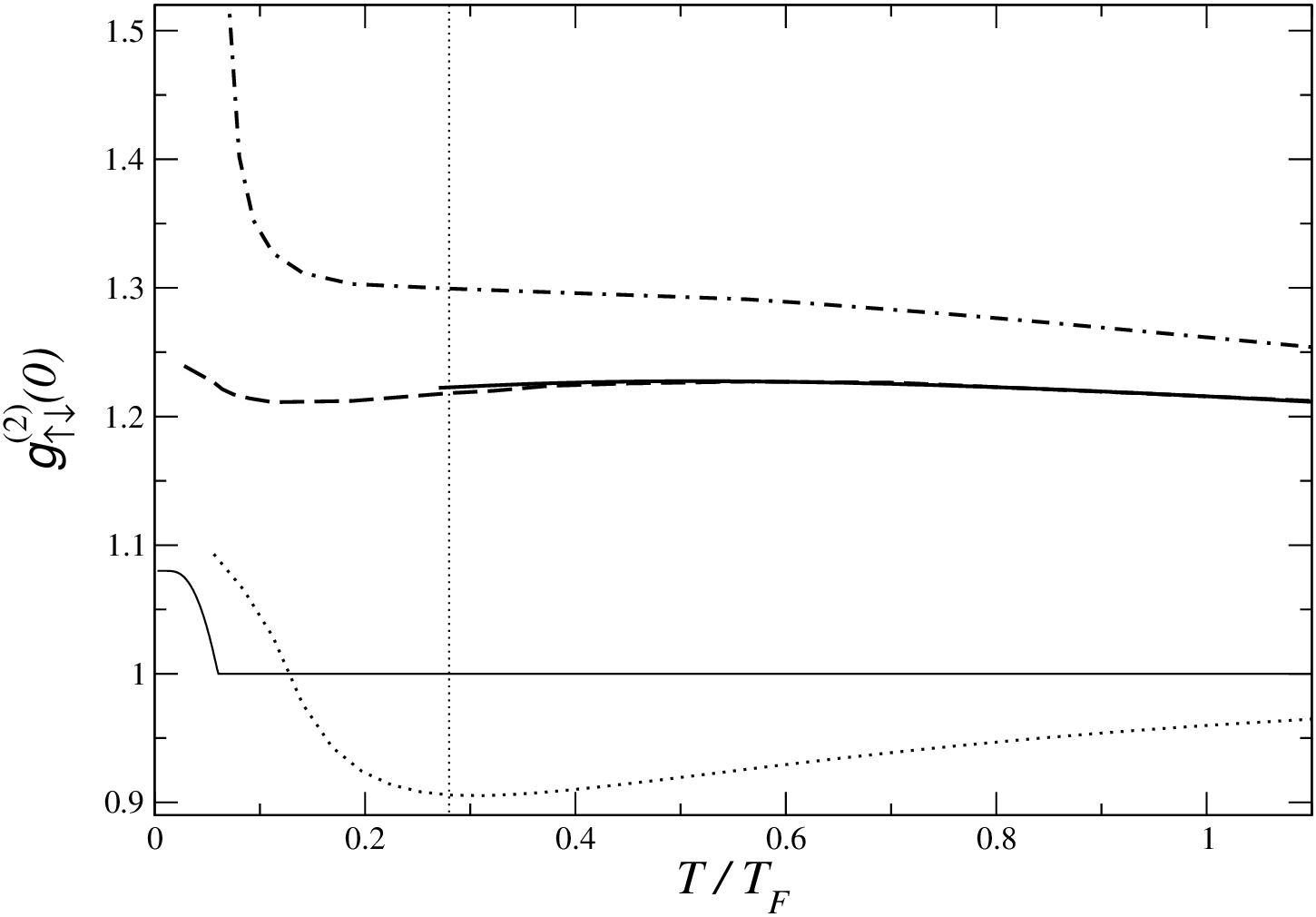}
\caption{
\label{fig:g2udGC_T}
Opposite spin density-density correlation function
  at $x=0$: $g\al{2}_{\updown}(0)$
  as a function of the temperature $T$ in the grand-canonical ensemble. 
 Solid line: perturbative result at
  order 2. Dashed line: density-density RPA. Dotted line:
 Nozi\`eres Schmitt-Rink theory. Dot-dashed line: $\Psi^\dagger
 \Psi^\dagger-\Psi\Psi$ RPA. Thin solid line: BCS theory.
The vertical line is an approximate lower bound on $T$ where the second order
perturbative theory is accurate; its position was determined by a comparison
to the Quantum Monte Carlo results of Fig.~\ref{fig:g2udC_T}.
Same system parameters as in fig.\ref{fig:g1uuMC}. The mean number of
 particles is fixed to $\langle N \rangle=12$.}
\end{center}
\end{figure}

To calculate $\delta\mu$ up to second order, we express the fact that the mean density
of spin up particles is fixed in $x=0$. As the system has translational and spin symmetry,
this is equivalent to fixing the mean total number of particles.
We therefore specialize Eq.(\ref{eq:exp_joli}) to
the case $O=\hat{\psi}^\dagger_{\uparrow}(0)\hat{\psi}_{\uparrow}(0)$
and obtain 
\footnote{A simpler expression for the denominator of the second order
term is $\partial_{\mu_0} \rho_{0\uparrow}$ where $\rho_{0\uparrow}$ is the density
in one spin component of the ideal Fermi gas with a chemical potential
$\mu_0$.}
\begin{equation}
\delta\mu = \frac{1}{2} g_0 \rho + g_0^2 \frac{\int_0^{\beta}d\tau_2 \int_0^{\tau_2}d\tau_1 \,
(dx)^2\sum_{x_1,x_2} P_{21} H_{21} (P_{21}H_{20}P_{10}-P_{20} H_{21} H_{10})}
{\int_0^\beta d\tau_1\, dx\sum_{x_1} P_{10} H_{10}} + O(g_0^3)
\end{equation}
where $dx$ is the spatial step of the grid, $\rho$ is the total density
and the following notations were introduced:
\begin{eqnarray}
P_{ij} &\equiv& \langle \hat{\psi}^\dagger_{\uparrow}(x_i,\tau_i)
\hat{\psi}_{\uparrow}(x_j,\tau_j)\rangle_0 \\ 
H_{ij} &\equiv& \langle \hat{\psi}_{\uparrow}(x_i,\tau_i)
\hat{\psi}^{\dagger}_{\uparrow}(x_j,\tau_j)\rangle_0  
\end{eqnarray}
for integers $i,j$ equal to $0,1$ or $2$ and with the convention $x_0=0,\tau_0=0$.
Note that the term of order $g_0$ in $\delta\mu$ coincides with the
Hartree-Fock mean field prediction. 

In a second step, we calculate $g^{(2)}_{\uparrow\downarrow}(0)$ by taking 
$O=\hat{\psi}^\dagger_{\uparrow}(0)\hat{\psi}_{\uparrow}(0) \hat{\psi}^\dagger_{\downarrow}(0)
\hat{\psi}_{\downarrow}(0)$. Eliminating $\delta\mu$ from the resulting expression
gives:
\begin{eqnarray}
(\rho/2)^2 g^{(2)}_{\uparrow\downarrow}(0) &=& (\rho/2)^2 -g_0 \int_0^{\beta} d\tau_1\,
dx \sum_{x_1} (P_{10}H_{10})^2  \nonumber \\
&+& g_0^2 \int_0^{\beta}d\tau_2 \int_0^{\tau_2}d\tau_1 \,
(dx)^2\sum_{x_1,x_2} (P_{21} H_{20} P_{10} - P_{20} H_{21} H_{10})^2 \nonumber \\
&+&O(g_0^3).
\label{eq:g2gc}
\end{eqnarray}
In the case of a negative coupling constant $g_0$, both the first and second order terms
are positive, leading to a spatial bunching of opposite spin particles, as expected
for attractive interaction.

From the comparison with the quantum Monte Carlo calculations in the canonical ensemble,
we know the temperature range over which the second order perturbative
expansion gives accurate predictions 
for $g^{(2)}_{\uparrow\downarrow}(0)$. For the grand canonical ensemble with the
same mean number of particles, we expect the same conclusion to apply.
We therefore use a numerical integration of Eq.(\ref{eq:g2gc}) 
as a test of existing approximate theories that will be reviewed in
sec.\ref{sec:Exist}. As is apparent in Fig.~\ref{fig:g2udGC_T}, the
density-density RPA is in very good agreement with the perturbative
result, whereas the $\Psi^\dagger\Psi^\dagger-\Psi\Psi$ RPA 
overestimates $g^{(2)}_{\uparrow\downarrow}(0)$ and the
Nozi\`eres-Schmitt-Rink prediction clearly underestimates it. 
At temperatures above the BCS critical temperature, the BCS theory
reduces to the mean-field Hartree-Fock theory which
gives for $g^{(2)}_{\uparrow\downarrow}(0)$  
simply the ideal Fermi gas result, $g^{(2)}_{\uparrow\downarrow}(0)=1$.

\section{Comparison with approximate theories}
\label{sec:Exist}

For the grand canonical ensemble, several approximate many-body
theories exist which can be used to obtain predictions for the
correlation functions of the interacting Fermi gas.
In the next subsections, some among the most famous ones are
discussed, namely the mean-field BCS theory~\cite{LandauCM,deGennes}, two
versions of the random-phase 
approximation (RPA) ~\cite{FetterWalecka,Mahan}, and the Nozi\`eres-Schmitt Rink (NSR)
theory~\cite{NSR} developed to study the BCS-BEC crossover in strongly
interacting gases.
A quantitative comparison with the prediction of a
grand canonical version of the perturbative expansion of section \ref{sec:Pert} will be
performed. 
In order for the comparison to be meaningful, the many-body theories
under investigation have been specialized to the specific case of the 
discrete lattice Hamiltonian \eq{Hamilt} with exacly the same
discretization parameters as used in the previous sections. 

\subsection{BCS theory}
\label{sec:BCS}
In the BCS theory, the equilibrium density matrix is determined 
in a self-consistent way from the mean-field quadratic Hamiltonian of
the grand canonical ensemble at a chemical potential $\mu$:
\begin{equation}
{\mathcal H}_{\rm BCS}=\sum_{k\sigma} \Big(\frac{\hbar^2 k^2}{2m}-\mu\Big)
\ahd_{k\sigma} \ah_{k\sigma} + 
g_0 \sum_{x\sigma} dx\,\rho_{-\sigma} \Psihd_\sigma(x)\Psih_\sigma(x)
-\sum_x \big(
\Psih_\uparrow(x) \Psih_\downarrow(x) \Delta^* +
\Psihd_\downarrow(x) \Psihd_\uparrow(x) \Delta\big)
\eqname{HamiltQ}
\end{equation}
where the mean density $\rho_\sigma$ in a given spin component and the
gap function $\Delta$ are defined as usual as:
\begin{eqnarray}
\rho_\sigma&=& \big\langle\Psihd_\sigma(x)\Psih_\sigma(x)\big\rangle \\
\Delta&=&-g_0\langle \Psih_\uparrow(x) \Psih_\downarrow(x) \big\rangle.
\end{eqnarray}
The quadratic Hamiltonian \eq{HamiltQ} is
easily diagonalized by a Bogoliubov transformation in the plane wave
basis:
\begin{equation}
{\mathcal H}_{\rm BCS}=\sum_{k\sigma} E_k \chd_{k\sigma}\ch_{k\sigma},
\end{equation}
where the $\ch$, $\chd$ operators satisfy Fermi anticommutation rules
and are related to the Fermi field operators by:
\begin{eqnarray}
\Psih_\downarrow(x)&=&\frac{1}{\sqrt{L}}\sum_k u_k\, e^{ikx} \; \ch_{k\downarrow}+ v_{k}\,
e^{ikx} \;
\chd_{-k,\uparrow} \\
\Psih_\uparrow(x)&=& \frac{1}{\sqrt{L}}\sum_k u_k\, e^{ikx} \; \ch_{k\uparrow}- v_{k}\,
e^{ikx} \; \chd_{-k\downarrow}
\eqname{BCS_field_exp}
\end{eqnarray}
where the positive coefficients $u_k$, $v_k$ of the Bogoliubov transformation
are defined by: 
\begin{equation}
u_k^2=1-v_k^2=\frac{1}{2}\left(1+\frac{\frac{\hbar^2 k^2}{2m}-{\tilde
    \mu}}{E_k}\right),
\end{equation}
the quasi-particle energies $E_k$ are given by:
\begin{equation}
E_k=\sqrt{\Delta^2+\left(\frac{\hbar^2 k^2}{2m}-{\tilde \mu}\right)^2}
\end{equation}
and the chemical potential is shifted as ${\tilde \mu}=\mu-g_0 \rho_\sigma$
so as to take into account the mean-field energy.
The self-consistency equation for the gap $\Delta$ is:
\begin{equation}
-\frac{g_0}{2L}\sum_k\frac{1-2f_k}{E_k}
=1, 
\eqname{gap}
\end{equation}
where the $f_k$ are the quasi-particle occupation numbers
$f_k=(e^{E_k/k_B T}+1)^{-1}$.
For high temperature $T>T_{\rm BCS}$, Eq.\eq{gap} has no solution, 
so that the system is in the {\em normal} phase $\Delta=0$ and the BCS
theory reduces to a Hartree-Fock theory.
At low temperature $T<T_{\rm BCS}$, the gap equation is solved for a
non-vanishing value of $\Delta$. This value grows as the temperature
decreases. 

\subsubsection{Calculation of correlation functions within the BCS
  theory}

The expansion of the field operator \eq{BCS_field_exp} in terms of
quasi-particle creation and destruction operators can be used to
obtain a prediction for the correlation functions.
For instance, the BCS prediction for the one-body correlation function
$g\al{1}_{\sigma\sigma}(x)$ is given by:
\begin{equation}
g\al{1}_{\sigma\sigma}(x)=\frac{1}{\rho_\sigma}\sum_k \frac{e^{-ikx} }{L}
\big[|u_k|^2\, f_k+|v_k|^2\, (1-f_k) \big].
\eqname{g1uuBCS}
\end{equation}
From Wick's theorem, the single-spin density-density correlation function
$g\al{2}_{\sigma\sigma}(x)$ is 
\begin{equation}
g\al{2}_{\sigma\sigma}(x)=1-\big|g\al{1}_{\sigma\sigma}(x)\big|^2.
\end{equation}
Both quantities are affected in a weak way by the attractive 
interactions and eventually by the appearance of a non-vanishing gap
$\Delta$. 

A richer physics can be found in the opposite spin density-density
correlation function $g\al{2}_{\uparrow\downarrow}(x)$. For this
quantity, the BCS theory predicts:
\begin{equation}
g\al{2}_{\upa\doa}(x)=1+\frac{1}{\rho_\upa \rho_\doa} \big| A(x) \big|^2,
\end{equation}
where the anomalous correlation function $A(x)$ is defined as:
\begin{equation}
A(x)=\big\langle\Psih_\doa(x)\Psih_\upa(0)\big\rangle=
\sum_k\frac{e^{ikx}}{L}u_k v_k \big(1-2f_k\big).
\end{equation}
As $\Delta=-g_0 A(0)$,
$g\al{2}_{\upa\doa}(0)$ has the simple expression:  
\begin{equation}
g\al{2}_{\upa\doa}(0)=1+\left|\frac{\Delta}{g_0\,\rho_\sigma}\right|^2. 
\end{equation}
For $T>T_{\rm BCS}$, this quantity is identically $1$, which means
that the BCS theory does not predict any correlation between the
densities in opposite spin states. These appear only for $T<T_{\rm
  BCS}$ as a consequence of the non-vanishing BCS gap.
As one can see in fig.\ref{fig:g2udGC_T}, this result is in qualitative
disagreement with the perturbative expansion which
gives a significant degree of correlation also for $T\simeq T_F\gg T_{\rm BCS}$.

The BCS prediction for the first-order pair coherence function
$g\al{1}_{\rm pair}(x)$ is:
\begin{equation}
g\al{1}_{\rm pair}(x)=
\frac{1}{\rho_\upa \rho_\doa g_0^2}\;|\Delta|^2+g_{\sigma\sigma}\al{1}(x)^2
\end{equation}
and is characterized by a short-ranged bump of spatial size of the
order of $\ell_F=1/k_F$, and a non-vanishing long-range limit.
As one can see in fig.\ref{fig:g1pairC_T}, the long distance behaviour
of $g\al{1}_{\rm pair}$ predicted by the BCS theory is in qualitative
agreement with the Monte Carlo predictions.

\subsection{Random Phase Approximation}

\subsubsection{Fluctuation-dissipation theorem}

A simple way of including the fluctuations around the mean-field is to
compute a response function within the mean-field theory and then
invoke the fluctuation-dissipation theory to obtain the corresponding
correlation function. 
In this subsection, we shall give a short review of the main results
of linear response theory that are required to obtain the correlation
functions of our interacting Fermi gas. 
A complete discussion of linear response theory and
fluctuation-dissipation theorem can be found
in~\cite{Levy_Magn,LandauCM}. 

Let $A$ and $B$ be two operators of a system characterized by a
time-independent Hamiltonian ${\mathcal H}$. For notational
simplicity, we assume that at equilibrium $\langle A \rangle_{\rm
  eq}=\langle B \rangle_{\rm eq}=0$. 
A weak perturbation of
the form:
\begin{equation}
{\mathcal H}_{\rm pert}=\epsilon(t)\,A^\dagger+\epsilon^*(t)\,A
\end{equation} 
is applied to the system and its effect on the observable $B$
recorded. At linear regime, this is summarized by the linear response
functions:
\begin{equation}
\langle B \rangle(t)=\int_{-\infty}^\infty
dt'\,\Big[
\chi_{BA}(t-t')\epsilon^*(t')+\chi_{BA^\dagger}(t-t')\epsilon(t')
\Big].
\end{equation}
The linear susceptibilities $\chi$ have the simple expression in terms
of commutators:
\begin{equation}
\chi_{BA}(t)=\frac{1}{i\hbar}\textrm{Tr}
\Big\{
\big[B(t),A\big]\rho_{\rm eq}
\Big\}\Theta(t),
\end{equation}
where $\rho_{\rm eq}=\frac{1}{\mathcal Z}e^{-\beta {\mathcal H}}$ is the
thermal equilibrium density matrix at $\beta=1/k_B T$, ${\mathcal
  Z}={\textrm Tr}[e^{-\beta{\mathcal H}}]$ is the partition function and
    $B(t)=e^{i{\mathcal H} t/\hbar} B e^{-i{\mathcal H}
      t/\hbar}$. 
As the Hamiltonian of the unperturbed system does not depend on time, the
Fourier transform of $\chi_{BA}(t)$ is the frequency-dependent
response function to a harmonic perturbation of frequency $\omega$:
\begin{equation}
{\tilde
  \chi}_{BA}(\omega)=\int_{-\infty}^\infty dt\,e^{i\omega
  t}\, \chi_{BA}(t)\,e^{-\eta t},
\end{equation}
where $\eta\rightarrow 0^+$ \footnote{A remark useful for the calculations to come
is to realize that ${\tilde \chi}_{BA}(\omega)$ has no delta
singularity in $\omega=0$.}
The correlation function ${\tilde S}_{BA}(\omega)$ is defined as:
\begin{equation}
{\tilde S}_{BA}(\omega)=\int_{-\infty}^\infty dt\,
e^{i\omega t}\,S_{BA}(t)=\int_{-\infty}^\infty dt\,
e^{i\omega t}\,
\big\langle B(t)\, A(0) \big\rangle.
\end{equation}

If the condition:
\begin{equation}
\textrm{Tr}[{\mathcal P}_{E_n}\,B\,{\mathcal P}_{E_m}\,A\,{\mathcal P}_{E_n}]\in {\mathbb R},
\eqname{Reality}
\end{equation}
holds for all the eigenenergies $E_n$ and $E_m$ of the Hamiltonian ${\mathcal
  H}$ where the projector ${\mathcal P}_{E}$ projects onto the
  eigenspace of energy $E$, then the fluctuation-dissipation
theorem holds in its most common form (Callen-Welton
theorem) relating the imaginary part of the response function 
$\textrm{Im}[{\tilde \chi}_{BA}(\omega)]$ to the correlation function
${\tilde S}_{BA}(\omega)$: 
\begin{equation}
\textrm{Im}[{\tilde \chi}_{BA}(\omega)]=-\frac{1}{2\hbar}{\tilde
  S}_{BA}(\omega)\big( 1-e^{-\beta \hbar \omega} \big).
\eqname{FD}
\end{equation}
It is easy to verify that the condition \eq{Reality} is verified if
$A^\dagger=B$ or, more generally, if $A^\dagger=S\,B\,S$, $S$ being an
arbitrary unitary operator such that $S^2=1$.

The fluctuation-dissipation theorem \eq{FD} implies that the
  correlation function ${\tilde  
  S}_{BA}(\omega)$ is fixed by the knowledge
  of $\textrm{Im}[{\tilde \chi}_{BA}(\omega)]$ modulo a delta
  distribution in $\omega=0$: 
\begin{equation}
{\tilde S}_{BA}(\omega)=-2\hbar \frac{\textrm{Im}[{\tilde
      \chi}_{BA}(\omega)]}{1-e^{-\beta \hbar
      \omega}}+2\pi\,C_{BA}\,\delta(\omega).
\eqname{FluctDiss}
\end{equation}
The constant $C_{BA}$ can be written as follows:
\begin{equation}
C_{BA}=-k_B T \big[\chi_{BA}^{\rm therm}-
\lim_{\omega\rightarrow 0}{\tilde \chi}_{BA}(\omega)\big]
\end{equation}
in terms of the thermodynamic (isothermal) susceptibility $\chi_{BA}^{\rm
  therm}$. As usual in thermodynamics, this is defined as the response
  on $B$ when the system is at thermal equilibrium in the presence of
  a weak and time-independent perturbation ${\mathcal H}_{\rm
  pert}=\epsilon^* A+\epsilon A^\dagger$:
\begin{equation}
\delta B^{\rm therm}=\epsilon^*\,\chi_{BA}^{\rm therm}+
\epsilon\;\chi^{\rm therm}_{BA^\dagger}
\end{equation}
resulting from the expansion to first order in $\epsilon$ of:
\begin{equation}
\delta B^{\rm therm}=\frac{\textrm{Tr}[B\, e^{-\beta ({\mathcal
        H}+\epsilon^* A+\epsilon A^\dagger  )}]}
{\textrm{Tr}[e^{-\beta ({\mathcal
        H}+\epsilon^* A+\epsilon A^\dagger  )}]}.
\end{equation}
Notice that
while the definition of $\chi^{\rm therm}_{BA}$ involves some implicit
coupling to a thermal reservoir at temperature $T$, ${\tilde
  \chi}_{BA}(\omega)$ is defined for an isolated system evolving under
the Hamiltonian ${\mathcal H}$. For this reason, the thermodynamical
susceptibility $\chi^{\rm therm}_{BA}$ and the static limit ${\tilde
  \chi}_{BA}(\omega\rightarrow 0)$ in general are not equal~\cite{Levy_Magn}.

From the microscopic expression of $C_{BA}$ in terms of the
eigenstates of ${\mathcal H}$ of energy $E_n$:
\begin{equation}
C_{BA}=-\frac{\beta}{\mathcal Z}
\sum_n e^{-\beta E_n}\,
\textrm{Tr}\big[{\mathcal P}_n\,B\,{\mathcal P}_n\,A\,{\mathcal P}_n\big],
\end{equation}
one concludes that $C_{BA}$ is not vanishing in the presence of
degeneracies between the eigenenergies of ${\mathcal H}$ or when the
diagonal matrix elements $\braopket{n}{A}{n}$ and $\braopket{n}{B}{n}$
are not vanishing.

The possibility of having a term in $\delta(\omega)$ in the
correlation function $S_{BA}$ is often neglected in
statistical mechanics textbooks, e.g.~\cite{LandauCM}. 
Although this is generally correct in the thermodynamical limit,
it may lead to incorrect results for the correlation functions of
finite systems. Examples of this issue are discussed in the next
section and in the Appendix \ref{app:delta(omega)}.

\subsubsection{Density-density RPA}

A prediction for the density-density correlation function of the Fermi
gas in the grand canonical ensemble can be obtained by applying
the general results of the previous subsection to the operator  
${\hat \rho}_\sigma(x)$ giving the particle density in the spin state
$\sigma$ at position $x$.
An approximate prediction for the density-density response functions
can be obtained  by linearizing the equations of the
mean-field theory discussed around the thermal equilibrium state. 
For historical reasons, this approximation scheme is usually called {\em
  random phase approximation} (RPA)~\cite{FetterWalecka,Mahan}.
For the sake of simplicity, we shall limit
ourselves to the case $T>T_{\rm BCS}$, regime in which the vanishing
of the anomalous averages considerably improves the physical
transparency of the formulas. 

As the system is spatially homogeneous with the same density
$\rho_\sigma$ in each spin-component, the different Fourier
components of the spatial density 
\begin{equation}
\delta{\hat \rho}_{k\sigma}=\frac{1}{\sqrt{L}}\sum_x\,dx\,e^{-ikx}\,
\big({\hat \rho}_\sigma(x)-\rho_\sigma\big)=
\frac{1}{\sqrt{L}}\sum_x\,dx\,e^{-ikx}\,\big(\Psihd_\sigma(x)\Psih_\sigma(x)-\rho_\sigma)
\end{equation}
are decoupled.
Taking $A=\delta{\hat \rho}_{k\sigma}$ and $B=\delta{\hat
  \rho}^\dagger_{k'\sigma'}$, the
frequency-dependent susceptibility matrix has the form:
\begin{equation}
{\tilde \chi}_{\sigma\sigma'}(k,k';\omega)=
{\tilde \chi}_{\sigma\sigma'}(k,\omega)\,\delta_{k,k'}.
\end{equation}
Because of the symmetry in the spin space,
${\tilde \chi}_{\upa\upa}={\tilde \chi}_{\doa\doa}$ and 
${\tilde \chi}_{\upa\doa}={\tilde \chi}_{\doa\upa}$,
so that the eigenvectors of the susceptibility matrix are the
symmetric and antisymmetric linear combination of the two spin
states. The corresponding eigenvalues are: 
\begin{equation}
{\tilde \chi}_\pm(k,\omega)={\tilde
  \chi}_{\upa\upa}(k,\omega)\pm{\tilde \chi}_{\upa\doa}(k,\omega). 
\end{equation}
Conversely, the susceptibility matrix ${\tilde
  \chi}_{\sigma\sigma'}$ in the $\sigma=\upa\doa$ basis 
is written as a function of the ${\tilde \chi}_\pm$ as:
\begin{equation}
{\tilde \chi}_{\sigma\sigma'}=\frac{1}{2}\left(
\begin{array}{ccc}
{\tilde \chi}_+ +{\tilde \chi}_- 
& & {\tilde \chi}_+ -{\tilde \chi}_- \\
{\tilde \chi}_+ -{\tilde \chi}_-
& & {\tilde \chi}_+ +{\tilde \chi}_-
\end{array}
\right)
\end{equation}

The RPA susceptibility of an interacting gas can be calculated from
the Hartree-Fock equation of motion~\cite{Blaizot}, and has the simple
expression: 
\begin{equation}
{\tilde \chi}_\pm(k,\omega)=
\frac{{\tilde \chi}_0(k,\omega)}{1\mp g_0\,{\tilde \chi}_0(k,\omega)},
\eqname{RPA_dens-dens}
\end{equation}
in terms of the susceptibility ${\tilde \chi}_0$ of a non-interacting,
one component Fermi gas at the same temperature and chemical
potential: 
\begin{equation}
{\tilde \chi}_0(k,\omega)=\frac{1}{L}
{\sum_q}\frac{f_q-f_{q+k}}{\hbar\omega+{\mathcal E}_q-{\mathcal E}_{q+k}+i\,0^+}.
\end{equation}
${\mathcal E}_q=\hbar^2 q^2/2 m$ are the energies of the single-particle states
and $f_q=(1+\exp[\beta({\mathcal E}_q-\mu)])^{-1}$ the corresponding Fermi occupation
factors. Notice that the quantity inside the sum vanishes for the
states $q$ such that ${\mathcal E}_q={\mathcal E}_{q+k}$.

An expression analogous to \eq{RPA_dens-dens}:
\begin{equation}
\chi^{\rm therm}_\pm(k)=
\frac{\chi^{\rm therm}_0(k)}{1\mp g_0\,\chi^{\rm therm}_0(k)},
\end{equation}
relates the thermodynamic susceptibilities $\chi^{\rm therm}_\pm(k)$
of the interacting gas to the thermodynamic susceptibility of the
non-interacting one-component gas: 
\begin{equation}
\chi^{\rm therm}_0(k)=-\frac{1}{L}\sum_q f_q (1-f_{q+k})\cdot \;\left\{
\begin{array}{ccc}
\frac{e^{\beta({\mathcal E}_q-{\mathcal E}_{q+k})}-1}{{\mathcal E}_q-{\mathcal E}_{q+k}} &
\textrm{if}&  {\mathcal E}_q\neq {\mathcal E}_{q+k}\\
\\
\beta & \textrm{if} &{\mathcal E}_q={\mathcal E}_{q+k}
\end{array}
\right.
\end{equation}

By applying the fluctuation-dissipation theorem in its form
\eq{FluctDiss} to the operators $B=\delta{\hat \rho}^\dagger_{k\sigma}$ and
$A=\delta{\hat \rho}_{k\sigma'}$, one can write the correlation function 
\begin{equation}
S_{\sigma\sigma'}(k,\omega)=\int_{-\infty}^\infty
dt\,e^{i\omega t}\,\big\langle
\delta{\hat \rho}^\dagger_{k\sigma}(t)\,\delta{\hat \rho}_{k\sigma'}(0)\big\rangle
\end{equation}
in terms of the imaginary part of ${\tilde
  \chi}_{\sigma\sigma'}(k,\omega)$ and the 
thermodynamic susceptibility $\chi_{\sigma\sigma'}^{\rm therm}(k)$:
\begin{equation}
S_{\sigma\sigma'}(k,\omega)=-2\hbar\,
\frac{\textrm{Im}[{\tilde \chi}_{\sigma\sigma'}(k,\omega)]}{1-e^{-\beta\hbar\omega}}
-2\pi\,k_B T\big[
\chi_{\sigma\sigma'}^{\rm therm}(k)-
\lim_{\omega\rightarrow 0}{\tilde \chi}_{\sigma\sigma'}(k,\omega)\big]\,\delta(\omega).
\eqname{S(k,omega)}
\end{equation}
The condition \eq{Reality} is here satisfied since $A$ and $B$ are 
 connected by $B^\dagger=S A S$ with $S$ respectively equal to the
 identity, if $\sigma=\sigma'$, or the spin-inversion operator $S$
 exchanging the spin components $\upa,\doa$ of all the particles,
 if $\sigma=-\sigma'$. 

Finally, the RPA prediction for the desired real-space, one-time
density-density correlation function can be found by inverse Fourier transform of
$S_{\sigma\sigma'}(k,\omega)$:
\begin{equation}
\big\langle {\hat \rho}_\sigma(x) {\hat \rho}_{\sigma'}(0)
\big\rangle=\rho_\sigma\rho_{\sigma'}+\frac{1}{L}\,\int_{-\infty}^\infty 
\frac{d\omega}{2\pi}\,\sum_k\,e^{-ikx}\,S_{\sigma\sigma'}(k,\omega).
\end{equation}

Corresponding predictions for the opposite spin density-density
correlation function at $x=0$ as a function of the temperature are
plotted in fig.\ref{fig:g2udGC_T}. Notice the excellent agreement of
the RPA prediction with the one of the perturbative expansion in
$g_0$ discussed in sec.\ref{sec:Pert}. 
In our finite system, the agreement strongly relies on the
correct inclusion of the $\delta(\omega)$ term in \eq{S(k,omega)}. An
explicit calculation of this issue for the non-interacting case is
presented in the Appendix \ref{app:delta(omega)}.

\subsubsection{$\Psi^\dagger\Psi^\dagger$-$\Psi\Psi$ RPA}

In the previous subsection, we have obtained a prediction for the
density-density correlation function $g\al{2}_{\sigma\sigma'}(x)$ of
an interacting Fermi gas by 
using the RPA density-density susceptibility and then invoking the
fluctuation-dissipation theorem.
In the present subsection, a similar approach is used to obtain the
pair coherence function $g\al{1}_{\rm pair}(x)$ at temperatures
higher than the BCS critical temperature $T_{\rm BCS}$ in the
grand canonical ensemble.

Consider the pair of operators:
\begin{eqnarray}
B&=&\Psihd_\doa(x)\Psihd_\upa(x) \\
A&=&\Psih_\upa(0)\Psih_\doa(0).
\end{eqnarray}
At thermal equilibrium both of them have a vanishing expectation
value. The correlation function:
\begin{equation}
\big\langle BA \big\rangle=\Big\langle \Psihd_\doa(x)\,
\Psihd_\upa(x)\,
\Psih_\upa(0)\,
\Psih_\doa(0)\Big\rangle=\rho_\upa \rho_\doa\,g\al{1}_{\rm pair}(x)
\end{equation}
can be evaluated from the susceptibility $\chi_{BA}$.

As in the previous subsection, we introduce the spatial Fourier
components as:
\begin{eqnarray}
A_k&=&\frac{1}{\sqrt{L}}\sum_x
dx\,e^{-ikx}\,\Psih_\upa(x)\Psih_\doa(x) \eqname{A_k} \\
B_k&=&\frac{1}{\sqrt{L}}\sum_x
dx\,e^{ikx}\,\Psihd_\doa(x)\Psihd_\upa(x); \eqname{B_k}
\end{eqnarray}
note the sign difference in the phase factors of \eq{A_k} and \eq{B_k}.
Thanks to the spatial homogeneity of the system, the susceptibility is 
diagonal in $k$-space. From the Hartree-Fock-Bogoliubov equation of
motion for the anomalous averages~\cite{Blaizot}, one can obtain the following simple
expression for the RPA susceptibility: 
\begin{equation}
{\tilde \chi}(k,\omega)=
\frac{{\tilde \chi}_0(k,\omega)}{1-g_0\,{\tilde \chi}_0(k,\omega)},
\end{equation}
where ${\tilde \chi}_0$ is defined as:
\begin{equation}
{\tilde
  \chi}_0(k,\omega)=\frac{1}{L}\sum_q\frac{2f_q-1}{\omega+{\mathcal
  E}_q+
{\mathcal E}_{q+k}+g_0 \rho-2\mu +i\,0^+}
\end{equation}
and describes the ideal gas response. As previously, ${\mathcal E}_q$ are the
energies of the single-particle states, $f_q$ the Fermi occupation
factors, and $\rho=\rho_\upa+\rho_\doa$ the total particle
density summed over both spin states.
Notice how $\chi(k=0,\omega=0)$ diverges when ${\tilde \chi}_0(k=0,\omega=0)$
tends to $1/g_0$. This is the signature of the approaching of the BCS
transition: the standard equation~\cite{LandauCM} for the BCS critical
temperature is in fact recovered if one imposes:
\begin{equation}
1=g_0\,\chi_0(k=0,\omega=0)=-\frac{g_0}{2L}\sum_q\frac{1-2f_q}{{\mathcal
    E}_q+
\frac{1}{2}g_0
  n-\mu}.
\end{equation}
As the chemical potential is a variable that can be continuously
varied, all degeneracies between states with different particle number
are accidental and occur only for discrete values of $\mu$. As $A$ and
$B$ have vanishing diagonal elements, the correction term in
$\delta(\omega)$ vanishes for all other values of $\mu$. As
$g\al{1}_{\rm pair}$ has a continuous dependance on $\mu$, there is no
need for calculating $\chi^{\rm therm}(k)$. We therefore have:
\begin{equation}
S_{BA}(k,\omega)=-2\hbar\frac{\textrm{Im}\big[{\tilde
      \chi}(k,\omega)\big]}{1-e^{-\beta\hbar\omega}} 
\end{equation}
and
\begin{equation}
g\al{1}_{\rm pair}(x)=\frac{1}{\rho_\upa \rho_\doa L}\int_{-\infty}^\infty
\frac{d\omega}{2\pi}\sum_k\,e^{-ikx}\,S_{BA}(k,\omega).
\end{equation}
The condition \eq{Reality} is here satisfied as $B_k^\dagger=A_k$.
As $g\al{1}_{\rm pair}(x=0)$ coincides with $g\al{2}_{\upa\doa}(x=0)$, we have included
in fig.\ref{fig:g2udGC_T} also the prediction of the present
$\Psi^\dagger\Psi^\dagger-\Psi\Psi$ RPA approach. 
The agreement with the perturbative expansion is less good than in the
case of the density-density RPA approach. 


\subsection{Nozi\`eres-Schmitt Rink approach}

In~\cite{NSR}, a non-perturbative calculation is performed for the
grand potential of a two component Fermi gas with attractive
interactions by a resummation of a certain class of diagrams.

Starting from Eq.~(20) of~\cite{NSR} which gives the grand potential
$\Omega$ in terms of an integral in the complex plane, we can deform the
integration contour and apply the residues formula to obtain the
following expression in terms of a sum for the case of a contact
interaction potential
\footnote{Our expression of $\Omega$ in terms of a Matsubara sum
  differs from the unnumbered equation between Eqs.~(19) and (20)
  of~\cite{NSR}. This is due to the omission by the authors 
  of~\cite{NSR} of the contribution to the contour integral of the
  half-circle of infinite 
  radius in the $\textrm{Re}\,\omega<0$ half plane.}:
\begin{equation}
\Omega(\mu,T;g_0)=\Omega_0(\mu,T)+k_B T \sum_{q,\omega_\nu}
\log[1-\chi(q,\omega_\nu;g_0)]-\frac{g_0}{2L} \sum_{k_1,k_2}
(1-f_{k_1}-f_{k_2}), 
\eqname{NSROmega}
\end{equation}
where $\omega_\nu=2i\pi\nu/\beta$ with $\nu$ integer ranging from
$-\infty$ to $+\infty$. The function $\chi$ is defined as:
\begin{equation}
\chi(q,\omega;g_0)=-\frac{g_0}{L}\sum_k\frac{1-f_k-f_{k+q}}
{{\mathcal E}_k+{\mathcal E}_{k+q}-2\mu-\omega},
\end{equation}
$f_q$ being the occupation number $f_q=\{1+
\exp[\beta({\mathcal E}_q-\mu)]\}^{-1}$ and
$\Omega_0$ being the grand potential for the ideal two-component Fermi
gas. 

This theory requires a self-consistent determination of $\mu$ from
$N=-\partial_\mu\Omega$, which we perform numerically. It gives
access to $g\al{2}_{\upa\doa}(0)$ 
thanks to the Hellmann-Feynman theorem~\cite{Lieb}:
\begin{equation}
\Big(\frac{\rho}{2}\Big)^2
g\al{2}_{\upa\doa}(0)=\frac{1}{L}\frac{\partial}{\partial g_0}\Omega.
\eqname{HellFey}
\end{equation}
The results are plotted in Fig.\ref{fig:g2udGC_T}. The poor agreement
with the perturbative expansion can be explained as follows.
Let us expand Eq.\eq{NSROmega} upto second order in $g_0$ at a given
$\mu$.
If one then replaces $\mu$ by its value in the NSR theory for the
density under consideration, one gets from Eq.\eq{HellFey} a prediction for
$g\al{2}_{\upa\doa}(0)$ upto first order that the can be compared to
the exact expansion Eq.\eq{g2gc}:
\begin{equation}
g\al{2}_{NSR}(0)-g_{\upa\doa}\al{2}(0)=\frac{\rho_0^2(\mu_{NSR})}{\rho^2}
-1+\textrm{O}(g_0^2) 
\eqname{g2NSRfin}
\end{equation}
Here $\rho_0(\mu)$ is the total density of the ideal two-component
Fermi gas for the chemical potential $\mu$. A first order expansion of
$\Omega$ is enough to obtain the chemical potential 
$\mu_{NSR}=\mu_0+g_0 \rho/2+\textrm{O}(g_0^2)$ in the
NSR theory and to conclude that $g\al{2}_{NSR}$
differs from the exact value by a term of the order of $g_0$ which has
the same sign as $g_0$.
This just because not all the second order diagrams for $\Omega$ have
been included in the resummation procedure
\footnote{The situation is more subtle in 3D: in the model of
  \cite{Randeria},
the bare coupling constant $g_0$ tends to zero when the cut-off energy
  tends to infinity, in which case the Hartree-Fock mean field term
in $\mu_{NSR}$ tends to zero. In this regime, a non-perturbative resummation
  procedure then seems unavoidable.}.

\section{Conclusions}
\label{sec:Conclu}

In the present paper, we have presented the result of extensive
Quantum Monte Carlo 
simulations for the static correlation functions of a one-dimensional
lattice model of
attractively interacting two component fermions.
The numerical results obtained by QMC have been compared to existing
approximate theories. Excellent agreement with the predictions of a
perturbative expansion in the interaction constant has been found, as well as
with the ones of the random phase approximation.

Although long-range order is destroyed by phase fluctuations in
one-dimensional systems in the thermodynamical limit, the finite size of the
system under consideration still allows for the identification of a crossover
to a condensed state at the temperature $T^*$ at which the first order
coherence length of the pairs becomes larger than the system size.

We have found that a significant degree of opposite spin
density-density correlations already exists at temperatures well above $T^*$
and is only slightly enhanced as the temperature goes below $T^*$.
This means that a measurement of the density-density correlation function
$g\al{2}_{\uparrow\downarrow}$ can not provide an unambiguous
signature of the onset of a condensed state in the gas. 
On the other hand, this could be provided by a measurement of the
second-order momentum space correlation function as suggested
in~\cite{Lukin}, or, even more directly, of the long-range behaviour
of the first-order pair coherence function $g\al{1}_{\rm  pair}(x)$.  
A non-vanishing limit of $g\al{1}_{\rm pair}(x)$ for large $x$ corresponds in
fact to the presence of a finite condensate fraction in both weak- (BCS) and
strong- (BEC) interaction regimes. 
A possible experimental scheme to measure $g\al{1}_{\rm pair}$ in
atomic Fermi systems by means of matter-wave interferometric techniques will be
the subject of a forthcoming publication.

\acknowledgments
Laboratoire Kastler Brossel is a Unit\'e de
Recherche de l'\'Ecole Normale Sup\'erieure et de l'Universit\'e Paris
6, associ\'ee au CNRS.
We acknowledge discussions with Ph. Chomaz, J. Dalibard, B. Derrida,
T. Jolic\oe ur, O. Juillet, A. Montina, C. Mora, A. Recati.

\appendix

\section{The Monte Carlo sampling algorithm}
\label{app:MC}

In the present appendix, we describe the numerical algorithm used for
the numerical simulations. A Monte Carlo technique has been used to
sample the probability distribution of the initial wavefunctions
$\phi_i\al{0}$ and the elementary noise terms $\xi^{(\alpha)}_j(x)$.

As the effective contributions of the different 
realizations to the observables involve the trace of the Hartree-Fock
ansatz $\sigma$, i.e. the scalar product between the two $N$-body
Nartree-Fock states, they can have enormously different values, so a direct
draw of the random variables $\phi_i\al{0}$ and $\xi^{(\alpha)}_j(x)$ would be
poorly efficient. 
An {\em importance sampling} scheme~\cite{NumRec} has therefore been
implemented, using the value of the modulus of the trace at the end
$\tau=\beta$ of the imaginary-time evolution as the {\em a 
  priori} probability distribution function $P_0$:
\begin{equation}
P_0\big[\big\{\phi_i\al{0},\xi^{(\alpha)}_j(x)\big\}\big]=\big|\textrm{Tr}[\sigma]\big|=
\big|\braket{\phi\al{2}_1(\beta)\ldots\phi\al{2}_N(\beta)}
{\phi\al{1}_1(\beta)\ldots\phi\al{1}_N(\beta)}\big|.
\end{equation}
In this way, the contributions of the different realizations to the
trace have the same absolute value, although their phases are still
random.

In order to sample $P_0$, a {\em Metropolis} scheme~\cite{KrauthNotes}
 has been implemented:
at each step a random move is proposed for both the wavefunctions
$\phi_i\al{0}$ and the elementary noises $\xi^{(\alpha)}_j(x)$.
For the first one, a rotation ${\mathcal R}_{{\mathbf n}\theta}$ in the one-body Hilbert
space is chosen with random rotation axis ${\mathbf n}$ and angle
$\theta$, and then is  
applied to all the orbitals $\phi_i\al{0}$:
\begin{equation}
\phi_i'\,\al{0}={\mathcal R}_{{\mathbf n}\theta}\;\phi_i\al{0}
\end{equation}
 This operation
rotates the hyperplane spanned by the set of orthonormal orbitals
$\{\phi_i\al{0}\}$ describing the initial Hartree-Fock state.
For what concerns the elementary noises $\xi^{(\alpha)}_j(x)$, each
of then is displaced to a new position $\xi'^{(\alpha)}_j(x)$ as follows:
\begin{equation}
\xi'^{(\alpha)}_j(x)=\sqrt{1-\eta^2}\,\xi^{(\alpha)}_j(x)+\eta\,b_j^{(\alpha)}(x),
\end{equation}
$b_j^{(\alpha)}(x)$ being independent, zero-mean, complex Gaussian variables such
that $\overline{b_j^{(\alpha)}(x)^2}=0$, $\overline{|b_j^{(\alpha)}(x)|^2}=1$. This kind of
random process is such that the resulting distribution of the $\xi$'s
is indeed a Gaussian with the required width.
The parameter
$\eta$ as well as the probability distribution for the random rotation angle
$\theta$ are free parameters which can be tuned to optimise the
efficiency of the simulation.
Denoting with $P_0\al{\rm in}$ and $P_0\al{\rm fin}$ the value of the a priori
probability for the configurations respectively before and after the
proposed move, this is accepted with a probability
$p=\textrm{min}[1, P_0\al{\rm fin} / P_0\al{\rm in} ]$.
As all configurations can be attained by the random motion and detailed
balance is verified, the stationary probability distribution of the
stochastic process is
indeed the desired one $P_0$. If a large enough number of moves is performed
between successive realizations, these can be considered to be
statistically independent.

\section{Expectation values for the calculation in the canonical ensemble}
\label{appen:expect}

The calculation of $g^{(2)}_{\uparrow\downarrow}(x_a)$ involves the
following expectation values:
\begin{equation}
I_n(\theta) = \langle \hat{\rho}_{\uparrow}(x_n,\tau_n)
\hat{\rho}_{\downarrow}(x_n,\tau_n) \ldots
\hat{\rho}_{\uparrow}(x_0,\tau_0)\hat{\rho}_{\downarrow}(x_a,\tau_a)\rangle_0(\theta),
\end{equation}
where $\hat{\rho}_{\sigma}(x,\tau) =
\hat{\psi}^\dagger_{\sigma}(x,\tau) \hat{\psi}_{\sigma}(x,\tau)$ and
$x_0=0$, $\tau_0=\tau_a=0$.  Introducing the notations for the
particle and hole correlation functions
\begin{eqnarray}
P_{ij} &\equiv& \langle \hat{\psi}^\dagger_{\uparrow}(x_i,\tau_i)
\hat{\psi}_{\uparrow}(x_j,\tau_j)\rangle_0(\theta) \\ H_{ij} &\equiv&
\langle \hat{\psi}_{\uparrow}(x_i,\tau_i)
\hat{\psi}^{\dagger}_{\uparrow}(x_j,\tau_j)\rangle_0(\theta)
\end{eqnarray}
where $i,j$ are integers from $0$ to $n$ or are equal to $a$, we find
\begin{eqnarray}
I_0 &=& P_{00}^2 \\ I_1 &=& (P_{00}^2+P_{1a} H_{1a}) (a\rightarrow 0)
\\ I_2 &=& (P_{00}^3+P_{21} H_{21} P_{00}+P_{1a} H_{1a} P_{00}+P_{2a}
H_{21} H_{1a} \nonumber \\ && -P_{21} H_{2a} P_{1a}+P_{2a} H_{2a}
P_{00})(a\rightarrow 0) \\ I_3 &=& (P_{32} H_{32} P_{00}^2+P_{32}
H_{32} P_{1a} H_{1a}-P_{32} H_{31} P_{2a} H_{1a}\nonumber \\ &&
-P_{32} H_{3a} P_{2a} P_{00}+ P_{32} H_{3a} P_{21} P_{1a}-P_{32}
H_{31} P_{21} P_{00}\nonumber \\ && +P_{2a} H_{21} H_{1a}
P_{00}+P_{2a} H_{2a} P_{00}^2 -P_{31} H_{3a} P_{1a} P_{00}\nonumber \\
&& +P_{3a} H_{31} H_{1a} P_{00}-P_{3a} H_{31} P_{21} H_{2a}+P_{31}
H_{32} H_{21} P_{00}\nonumber \\ && -P_{31} H_{32} H_{2a} P_{1a}
+P_{3a} H_{3a} P_{00}^2+P_{3a} H_{3a} P_{21} H_{21} \nonumber \\ &&
+P_{31} H_{31} P_{2a} H_{2a}+P_{31} H_{31} P_{00}^2+P_{00}^4 +P_{1a}
H_{1a} P_{00}^2\nonumber \\ && +P_{21} H_{21} P_{00}^2-H_{2a} P_{21}
P_{1a} P_{00}+P_{3a} H_{32} H_{21} H_{1a}\nonumber \\ && + P_{3a}
H_{32} H_{2a} P_{00}-P_{31} H_{3a} P_{2a} H_{21}) (a\rightarrow 0)
\end{eqnarray}
The property that each of these expressions is a product of two
similar factors, the second one deduced from the first one by the
replacement $a\rightarrow 0$, originates from the fact that the ideal
Fermi gas in the grand canonical ensemble consists of two equivalent
and independent spin components.

Similarly the calculation of $g_{\rm pair}^{(1)}(x_a)$ involves the
expectation values
\begin{equation}
J_n(\theta) = \langle \hat{\rho}_{\uparrow}(x_n,\tau_n)
\hat{\rho}_{\downarrow}(x_n,\tau_n) \ldots
\hat{\psi}^\dagger_{\uparrow}(x_a)\hat{\psi}^\dagger_{\downarrow}(x_a)
\hat{\psi}_{\downarrow}(0)\hat{\psi}_{\uparrow}(0)\rangle.
\end{equation}
Using the previous notations we then obtain
\begin{eqnarray}
J_0 &=& P_{a0}^2 \\ J_1 &=& (P_{00}P_{a0}+P_{10}H_{1a})^2 \\ J_2 &=&
(P_{20} H_{21} H_{1a}+H_{2a} P_{20} P_{00}+P_{00}^2 P_{a0} \nonumber
\\ && +P_{10} H_{1a} P_{00}+P_{a0} P_{21} H_{21}-P_{21} H_{2a}
P_{10})^2 \\ J_3 &=& (P_{31} H_{31} P_{a0} P_{00}+P_{00}^3
P_{a0}-P_{31} H_{3a} P_{10} P_{00}+P_{00}^2 P_{10} H_{1a} \nonumber \\
&&+P_{20} H_{2a} P_{00}^2+P_{20} H_{21} H_{1a} P_{00}-P_{21} H_{2a}
P_{10} P_{00} \nonumber\\ &&+P_{21} H_{21} P_{00} P_{a0}+P_{30} H_{31}
H_{1a} P_{00}+P_{30} H_{32} H_{2a} P_{00} \nonumber\\ &&+P_{32} H_{32}
P_{a0} P_{00}-P_{32} H_{3a} P_{20} P_{00}+P_{31} H_{32} H_{21} P_{a0}
\nonumber\\ &&-P_{32} H_{31} P_{21} P_{a0}+P_{32} H_{32} P_{10}
H_{1a}-P_{32} H_{31} P_{20} H_{1a} \nonumber\\ &&+P_{32} H_{3a} P_{21}
P_{10}+P_{30} H_{32} H_{21} H_{1a}-P_{30} H_{31} H_{2a} P_{21}
\nonumber\\ &&+P_{30} H_{3a} P_{21} H_{21}-P_{31} H_{32} H_{2a}
P_{10}+P_{31} H_{31} P_{20} H_{2a} \nonumber\\ && -P_{31} H_{3a}
H_{21} P_{20}+P_{30} H_{3a} P_{00}^2)^2.
\end{eqnarray}
The fact that these expressions are squares is again due to the
existence of two independent and equivalent spin components.

\section{Density fluctuations of an ideal Fermi gas: physical meaning of
  the $\delta(\omega)$ term}
\label{app:delta(omega)}

The density-density correlation function for a one-component
non-interacting Fermi gas in the grand canonical ensemble at inverse
temperature $\beta=1/k_B T$ and chemical potential $\mu$ can be
calculated from Wick's theorem: 
\begin{equation}
G\al{2}(x)=\big\langle {\hat \rho}(x) \, {\hat \rho}(0)
\big\rangle-\rho^2=\frac{1}{L^2}\sum_{qk}e^{ikx} f_q(1-f_{q+k}),
\eqname{WickApp}
\end{equation}
where ${\hat \rho}(x)$ is the operator giving the density at point $x$.
A calculation based on the fluctuation-dissipation theorem neglecting
the term $2\pi C_{BA}\,\delta(\omega)$ term in \eq{FluctDiss} would give:
\begin{equation}
G\al{2}(x)\stackrel{?}{=}\frac{1}{L^2}{\sum_{qk}}^{\neq} e^{ikx}
\frac{f_q-f_{q+k}}{1-e^{\beta({\mathcal E}_{q}-{\mathcal E}_{q+k})}}
\eqname{FDApp}
\end{equation}
where the sum $\sum^{\neq}$ has to be performed over the pair of states such that
${\mathcal E}_q\neq {\mathcal E}_{q+k}$.
Using the relation:
\begin{equation}
\frac{1-f_q}{f_q}=e^{\beta({\mathcal E}_q-\mu)},
\end{equation}
one can see that expression \eq{FDApp} does not coincide with
\eq{WickApp} because of the missing contribution of the pairs such that
${\mathcal E}_q={\mathcal E}_{q+k}$.

Inclusion of the term proportional to $\delta(\omega)$ in
\eq{FluctDiss} fixes the problem, since it exactly provides the
missing contribution:
\begin{equation}
\Delta{G}\al{2}(x)=\frac{1}{L^2}{\sum_{qk}}^{=}e^{ikx}
f_q(1-f_{q+k}),
\eqname{FDApp2}
\end{equation}
where the sum $\sum^{=}$ has to be performed over the pairs of states
such that ${\mathcal E}_q={\mathcal E}_{q+k}$. The physical meaning of the
$k=0$ term which contains the contribution of the diagonal matrix
elements of the perturbation is transparent: it keeps track of the
total particle number fluctuations of the grand canonical ensemble. 
In our spatially homogeneous system, degenerate pairs of states for $k=-2q$ are
also present. 

It is apparent from \eq{FDApp2} that the contribution of the
$\Delta{G}\al{2}$ correction term tends to zero in the
thermodynamical limit $L\rightarrow \infty$.


\begin{thebibliography}{99}

\bibitem{FermiDegen} B. De Marco  and D. S. Jin, Science {\bf 285},
  1703 (1999); 
A. C. Truscott, K. E. Strecker, W. I. McAlexander,
  G. B. Partridge, R. G. Hulet, Science {\bf
  291}, 2570 (2001);  
F. Schreck, L. Khaykovich, K. L. Corwin,
  G. Ferrari, T. Bourdel, J. Cubizolles, and C. Salomon,
  Phys. Rev. Lett. {\bf 87}, 080403 (2001); 
S. R. Granade, M. E. Gehm,
  K. M. O'Hara, and J. E. Thomas, Phys. Rev. Lett. {\bf 88} 120405 (2002);
    Z. Hadzibabic, C. A. Stan, K. Dieckmann, S. Gupta,
  M. W. Zwierlein, A. G\"orlitz, and W. Ketterle, Phys. Rev. Lett. {\bf
  88}, 160401 (2002). 

\bibitem{Molecules} J. Cubizolles, T. Bourdel,
  S. J. J. M. F. Kokkelmans, G. V. Shlyapnikov, and C. Salomon,
  Phys. Rev. Lett. {\bf 91}, 240401 (2003); 
C. A. Regal,  C. Ticknor, J. L. Bohn, D. S. Jin, Nature {\bf 424},
  47 (2003); S. Jochim, M. Bartenstein, A. Altmeyer, G. Hendl,
  C. Chin, J. Hecker Denschlag, and R. Grimm, Phys. Rev. Lett. {\bf
  91}, 240402 (2003).


\bibitem{MolecBEC} M. Greiner, C. A. Regal, D. S. Jin, Nature {\bf
  426}, 537 (2003); S. Jochim, M. Bartenstein, A. Altmeyer, G. Hendl,
  S. Riedl, C. Chin, J. Hecker Denschlag, and R. Grimm, Science {\bf
  302}, 2101 (2003)

\bibitem{CrossoverExp} M. Bartenstein,
  A. Altmeyer, S. Riedl, S. Jochim, C. Chin, J. Hecker Denschlag,
  R. Grimm, preprint cond-mat/0401109;
T. Bourdel, L. Khaykovich, J. Cubizolles,
  J. Zhang, F. Chevy, M. Teichmann, L. Tarruell,
  S.J.J.M.F. Kokkelmans, C. Salomon, preprint cond-mat/0403091.

\bibitem{AtomicBCS} C. A. Regal, M. Greiner, and D. S. Jin,
  Phys. Rev. Lett. {\bf 92}, 040403 (2004); M. W. Zwierlein,
  C. A. Stan, C. H. Schunck, S. M. F. Raupach, A. J. Kerman, and
  W. Ketterle, Phys. Rev. Lett. {\bf 92}, 120403 (2004).


\bibitem{NSR} P. Nozi\`eres and S. Schmitt-Rink,
  J. Low. Temp. Phys. {\bf 59}, 195 (1985)

\bibitem{Randeria}  M. Randeria, in {\em Bose-Einstein
  Condensation}, edited by 
  A. Griffin, D. W. Snoke, and S. Stringari (Cambridge, New York,
  1995), p. 355.

\bibitem{CrossoverTheory}
M. Holland, S. J. J. M. F. Kokkelmans, M. L. Chiofalo, and R. Walser,
  Phys. Rev. Lett. {\bf 87}, 120406 (2001);
Y. Ohashi and A. Griffin,  Phys. Rev. Lett. {\bf 89}, 130402 (2002);
J. Carlson, S.-Y. Chang, V. R. Pandharipande, and K. E. Schmidt,
  Phys. Rev. Lett. {\bf 91}, 050401 (2003).

\bibitem{Chomaz} O. Juillet, Ph. Chomaz, D. Lacroix, and
F. Gulminelli, Phys. Rev. Lett. {\bf 88}, 142503 (2002).

\bibitem{LandauCM} L. D. Landau, E. M. Lifshitz, and L. P. Pitaevskii,
  {\em Statistical Physics}, Vols.1 and 2, Pergamon Press, Oxford, 1980.  

\bibitem{deGennes} P.-G. de Gennes, {\em Superconductivity of metals
  and alloys}, Addison-Wesley, Redwood City, 1989.

\bibitem{FetterWalecka} A. L. Fetter and J. D. Walecka, {\em Quantum theory of
  many-particle systems}, McGraw-Hill, 1971.

\bibitem{Mahan} G. Mahan, {\em Many-particle physics}, Plenum Press,
  New York, 1981.

\bibitem{g2ud_obs} J. Ruostekoski, Phys. Rev. A {\bf 60}, 1775R (1999);
  F. Weig and W. Zwerger, Europhys. Lett. {\bf 49},  282 (2000).

\bibitem{MoraThese} C. Mora, Ph.D. thesis, unpublished (2004).

\bibitem{YvanHouches} Y. Castin, Lecture Notes of the 2003 Les Houches School
on {\em Quantum Gases in Low Dimensions}, EDP Sciences (2004).

\bibitem{Jacobi} see, e.g.: A. C. Aitken {\em Determinants and
  Matrices}, Oliver and Boyd, Edinburgh, 1956; F. R. Gantmacher {\em Theory of
  Matrices}, Nauka, Moscow, 1967.

\bibitem{Chomaz2} O. Juillet, F. Gulminelli, Ph. Chomaz, preprint cond-mat/0311437.

\bibitem{FermionMC} for a review, see: R. R. dos Santos,
  Braz. J. Phys. {\bf 33}, 36 (2003) and references therein. 

\bibitem{SpinGap} M. Randeria, N. Trivedi, A. Moreo, and
  R. T. Scalettar, Phys. Rev. Lett. {\bf 69}, 2001 (1992).

\bibitem{T_cMC1} K. Kuroki, H. Aoki, Phys. Rev. B {\bf 56}, 14287
  (1997); K. Kuroki, H. Aoki, J. Phys. Soc. Japan {\bf 67}, 1533 (1998).

\bibitem{T_cMC2} T. Paiva, R. R. dos Santos, R. T. Scalettar,
  and P. J. H. Denteneer, preprint cond-mat/0403397.

\bibitem{T_c3d} R. R. dos Santos, Phys. Rev. B {\bf 50}, 635 (1994).

\bibitem{1Dg1pair} R. A. Ferrell, Phys. Rev. Lett. {\bf 13}, 330
  (1964); S. Traven, Phys. Rev. Lett. {\bf 73}, 3451 (1994).

\bibitem{Lukin}  E. Altman, E. Demler, M. D. Lukin,
preprint cond-mat/0306226

\bibitem{Levy_Magn} L.-P. L\'evy, {\em Magn\'etisme et
  supraconductivit\'e}, Inter\'Editions / CNRS \'Editions, Paris,
  1997, chapter 8.

\bibitem{Blaizot} J.-P. Blaizot and G. Ripka, {\em Quantum theory of finite
  systems}, MIT Press, 1986.

\bibitem{Lieb} E. H. Lieb and W. Liniger, Phys. Rev. {\bf 130}, 1605
  (1963).

\bibitem{NumRec} W. H. Press, S. A. Teukolsky, W. T. Vetterling, and B. P.
Flannery, {\rm Numerical Recipes} (Cambridge University Press,
Cambridge, 1988).

\bibitem{KrauthNotes}  W. Krauth, {\em Introduction to Monte Carlo
  Algorithms}, in "Advances in Computer Simulation" (J. Kertesz and
  I. Kondor, eds) Lecture Notes in Physics (Springer Verlag, 1998).

\end{thebibliography}
\end{document}